\documentclass{ar-1col}
\usepackage{url}
\usepackage[numbers]{natbib}
\usepackage{epsfig}
\usepackage{comment} 
\usepackage[top=25truemm,bottom=25truemm,left=40truemm,right=15truemm]{geometry}
\usepackage{multirow}

\newcommand{\PRF}{\textcolor{black}} 

\jname{Xxxx. Xxx. Xxx. Xxx.}
\jvol{AA}
\jyear{YYYY}
\doi{10.1146/((please add article doi))}

\begin{document}

\markboth{Kato et al.}{Review of pre-SN neutrinos}

\title{Theoretical prediction of pre-supernova neutrinos and their detection}

\author{C. Kato$^1$, K. Ishidoshiro$^2$ and T. Yoshida$^3$
\affil{$^1$Department of Aerospace Engineering, Tohoku University, Sendai 980-8579, Japan; e-mail:  chinami.kato.e8@tohoku.ac.jp}
\affil{$^2$Research Center for Neutrino Science, Tohoku University, Sendai 980-8578, Japan; e-mail: koji@awa.tohoku.ac.jp}
\affil{$^3$Department of Astronomy, Graduate School of Science, The University of Tokyo, Tokyo 113-0033, Japan; e-mail: tyoshida@astron.s.u-tokyo.ac.jp}}

\begin{abstract}
Almost 30 years have passed since the successful detection of supernova neutrinos from SN 1987A.
In the last decades, remarkable progress has been made in neutrino detection technique, through which it may be possible to detect neutrinos from a new source, pre-supernova (pre-SN) neutrinos.
They are emitted from a massive star prior to core bounce.
Because neutrinos escape from the core freely, they carry information about the stellar physics directly.
Pre-SN neutrinos may play an important role in verifying our understanding of stellar evolution for massive stars.
Observations of pre-SN neutrinos, moreover, may serve as an alarm regarding a supernova explosion a few days in advance if the progenitor is located in our vicinity, enabling us to observe the next galactic supernova.
In this review, we summarize the current status of pre-SN neutrino studies from both of the theoretical and observational points of view.
\end{abstract}


\maketitle

\tableofcontents


\section{INTRODUCTION}
     A big breakthrough came in neutrino observations in 1987.
     A massive star exploded in the Large Magellanic Cloud, which is located at $\sim$ 50 kpc from the Earth.
     Such a big explosion at the end of \PRF{the life of} a massive star is called \PRF{a} "supernova".
     KAMIOKANDE-II, Irvine-Michigan-Brookhaven, and Baksan, which were neutrino detectors running at that time, detected \textcolor{black}{$\sim$} 11, 8 and 5 neutrinos, respectively, accompanying "SN 1987A" \citep{hirata1987, Bionta1987,Alekseev:1987ej}. 
     From their data, the average energy of electron anti-neutrinos ($\bar{\nu}_e$'s) was found to be  $\sim$ 15 MeV\PRF{,}
     and \PRF{the energy released was} $\sim 8\times10^{52}$ ergs \cite{hirata1987}. 
     These outcomes were consistent with \PRF{the} neutrino heating mechanism, which is one of the most favored explosion mechanisms \citep{sato1987}. 
     \PRF{This discovery advanced} our understanding of supernova explosions.
     
     In the last few decades, remarkable progress has been made \PRF{in neutrino detection techniques}.
     Neutrino detectors with large volumes and low-energy thresholds for neutrinos \PRF{have} been running such as Super-Kamiokande (SK) and Kamioka Liquid-scintillator Antineutrino Detector (KamLAND).
     A further scale-up will be realized \PRF{in} future detectors \PRF{such as} Hyper-Kamiokande (HK) and Jiangmen Underground Neutrino Observatory (JUNO). 
     It is worth mentioning the low background techniques that use a delayed coincidence (DC) for inverse-$\beta$ decay (IBD) to confirm \PRF{the} events \PRF{via} prompt and delayed signals with the \PRF{temporal} and spatial correlations.
     The SK-Gd project (Super-Kamiokande with gadolinium) \PRF{uses} such techniques \cite{simpson2019}.
     \PRF{Past and current} neutrino detectors are mainly sensitive to $\bar{\nu}_e$'s, whereas \PRF{newer} detectors \PRF{with the ability} to detect other flavors may become available in the near future.
     For the detection of electron neutrinos ($\nu_e$'s), 
     the Helium and Lead Observatory (HALO) is running at SNOLAB based on neutrino reactions on lead~\cite{ZUBER2015233}, and 
     Deep Underground Neutrino Observatory (DUNE) is currently under construction at Sanford Underground Research Facility (SURF) \cite{dune2016}.
     The detection of heavy-lepton neutrinos ($\nu_x$'s) \PRF{can be realized} by the observation of \textcolor{black}{coherent elastic neutrino-nucleus scattering (CE$\nu$NS)}, as in the COHERENT experiment at the Spallation Neutron Source \cite{akimov2017}. 
     Sensitivity of dark matter detectors to $\nu_x$'s via \textcolor{black}{CE$\nu$NS} is discussed \PRF{in} References \cite{ABE201751,Rafael2016,KOZYNETS201925}. 
     Multi-flavor studies for astrophysical neutrino sources will be available soon.
     
     Such developments may enable the detection of neutrinos from new sources. 
     \PRF{"Pre-supernova (pre-SN) neutrinos", which are mainly emitted from a core of a massive star but prior to core bounce, are one candidate.}
     Massive stars are thought to be the progenitors of core-collapse supernovae (CCSNe) \cite[e.g.,][]{Smartt2015}.
     They evolve \PRF{with} synthesizing heavier elements \PRF{like} He, C, Ne, O, Si and Fe via nuclear burnings.
     \PRF{Efficient neutrino emission occurs} by thermal emission processes \PRF{in a stellar core} with high temperature and density after a carbon core is formed.
     The nuclear weak processes [i.e., electron capture (EC), in which an electron is captured by a nucleus and a nuclear proton is changed to a neutron and $\beta$ processes] enhance neutrino emission after iron-group elements are synthesized.
     Neutrinos cool down the core and have important roles in its thermal evolution.
     
     The possibility of pre-SN neutrino detection was \PRF{first proposed} by Odrzywolek {\it et al.} \cite{odrzywolek2004}. 
     They calculated the luminosities and spectra of the neutrinos emitted via electron-positron pair annihilation for a 20 $M_\odot$ progenitor model and estimated the detection events for \PRF{six} neutrino detectors. 
     Assuming that the distance to a supernova is 1 kpc, the event numbers would be 41 for SK and, 4 for KamLAND.
     This \PRF{work}, a pioneer of qualitative calculations for pre-SN $\bar{\nu}_e$'s, enables our discussions of "pre-SN neutrino astronomy".
     
     Recent studies \PRF{have employed} state-of-the-art stellar evolution models \PRF{to} calculate the time evolution of \PRF{the} number luminosities and spectra for pre-SN neutrinos \cite{Kato2015,Yoshida2016,Kato2017,Patton2017b}.
     In addition, a variety of neutrino reactions are included in their calculations.
     Another thermal process---\PRF{namely} a plasmon decay---becomes dominant in the late \PRF{phases} of some progenitors \cite{Kato2015}.
     It also has been pointed out that neutrino emissions via nuclear weak interactions may dominate over the thermal processes just prior to collapse \cite{Kato2017, Patton2017b, odrzywolek2009}.

     Pre-SN neutrino astronomy \PRF{may} clarify current uncertainties in massive stars and supernovae.
     \PRF{First}, the detection of pre-SN neutrinos \PRF{can be} a direct observation of stellar interiors.
     The investigation of stellar evolutions has a long history, and a standard model has almost been established. 
     \PRF{Stellar physics}, however, still has many uncertainties.
     Pre-SN neutrinos may be one of the most promising tools to look in
     \PRF{stellar physics} because they can propagate through stars freely and \PRF{can} be detected without losing the thermal information of the core.
     \PRF{It has been} reported that pre-SN neutrino observations will make it possible to distinguish between the two types of CCSN-progenitors---iron core collapse supernovae (FeCCSNe) and electron capture supernovae (ECSNe)---\cite{Kato2015} and to impose restrictions on convective properties associated with oxygen-shell (O-shell) and silicon-shell (Si-shell) burnings \cite{Yoshida2016}.
     \PRF{Second}, the observation of pre-SN neutrinos serves as an alarm (SN alarm) regarding the subsequent explosion, and makes \PRF{it} possible to observe the next galactic supernova.
     The neutrinos may be detected a few days before the explosion, if the progenitor is located \PRF{in} our vicinity ($<$ 1 kpc) \citep{Kato2017,Raj2019}.
     It is supposed that galactic supernovae occur once every \PRF{few hundreds years.} Thus, the SN alarm has \PRF{an} important role \PRF{in helping us understand} the mechanism of supernova explosions.
     We believe that the detection of pre-SN neutrinos \PRF{in the future} will \PRF{have an} impact as \PRF{big} as that of the historical neutrino events at SN 1987A.
     
     In this review, we introduce pre-SN neutrinos \PRF{from} both the theoretical and observational points of view.
     We \PRF{first} summarize \PRF{the evolutions of massive stars} and neutrino emission in Section~\ref{sec2}.
     We then show the current status of neutrino detectors and discuss \PRF{a pre-SN neutrino alarm} in Section~\ref{sec3}.
     Section~\ref{sec3} also provide the expected numbers of pre-SN neutrino events \PRF{in} current and future detectors.
     In Section~\ref{sec4}, we discuss what can \PRF{be learnt} from future observations of pre-SN neutrinos; we focus on the distinction of supernova progenitors, the restrictions on shell burnings, and the determination of \PRF{the} neutrino mass ordering.
     Finally, we summarize the review and mention the future prospects of pre-SN neutrino studies in Section~\ref{sec5}.



\section{THEORY OF PRE-SUPERNOVA NEUTRINOS} \label{sec2}

     \subsection{Massive Star Evolution} \label{stellarevo}
\begin{figure}
 \centering
 \vspace*{-2.6cm}
 \includegraphics[scale=0.38,angle=270]{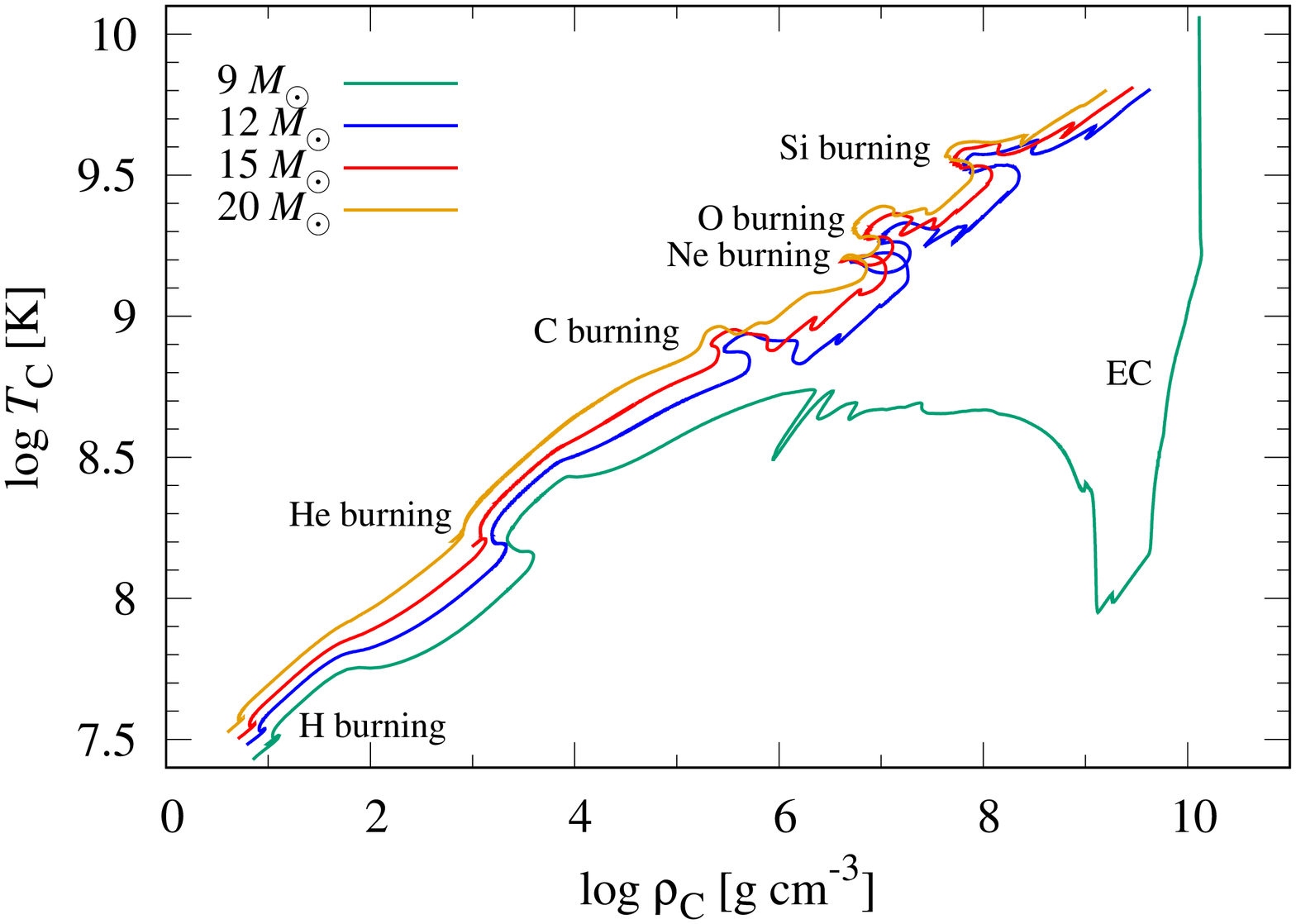}
  \vspace*{-3.5cm}
 \caption{Evolutionary paths in the central density ($\rho_{\rm C}$) and temperature ($T_{\rm C}$) plane. The figure shows 12 (blue), 15 (red) and 20 $M_\odot$ (orange) progenitors of FeCCSNe and \PRF{a} 9 $M_\odot$ progenitor (green) of ECSNe.}
 \label{ref:rhoc_tc}
\end{figure}

    Massive stars---that is, stars \PRF{with masses} heavier than $\sim$ 8 $M_\odot$ at the zero-age main sequence (ZAMS)---start their lives with hydrogen burning and end by a dynamical collapse of the central core\PRF{s} \cite[e.g.,][]{Woosley2002,Langer2012}.
    There are two types of progenitors that produce CCSNe. 
    In most cases, the stellar core is mainly composed of iron (Fe core)\PRF{,} and its collapse leads to an FeCCSN\PRF{. In} other cases, which \PRF{account for} $\sim$ 5\% of all CCSNe according to a recent study \citep{doherty2017}, \PRF{the} core is composed of oxygen and neon (ONe core)\PRF{,} and an ECSN occurs at the end.
    The initial mass \PRF{of} main sequence \PRF{stars} is the primary factor determining which \PRF{state} is obtained in the end: 
    Less massive stars (mass less than $\sim 9.5\ M_\odot$) will lead to ECSNe\PRF{,} and more massive stars will produce FeCCSNe \citep{umeda2012,jones2013,Takahashi2013,Doherty2015}.
    We explain \PRF{the} evolution sequence of a 15 $M_\odot$ model \cite{Yoshida2016, Kato2017} as an example of \PRF{an} FeCCSNe-progenitor.
    We then focus on the difference\PRF{s} in evolution between the two types of progenitors, since they pass through quite different paths\PRF{, as shown} in Figure~\ref{ref:rhoc_tc}.
    


    In the main sequence of massive stars, hydrogen is burned into helium via CNO cycles.
    The ignition temperature for hydrogen burning is $\sim 4 \times 10^7$ K\PRF{,} and it lasts for $\sim$ 10$^{7}$ yr for the 15~ $M_\odot$ model.
    \PRF{Stars} in this phase are observed as blue supergiants.
    After the central hydrogen is exhausted, the core contracts and has an isothermal structure.
    Its surrounding hydrogen-shell burning makes the pressure gradient larger at the bottom of the hydrogen shell and extends the hydrogen envelope.
    Such a star is called a red supergiant (RSG).

    \PRF{Helium ignites at a core temperature of} $\sim$ $1.5 \times 10^8$ K.
    This process proceeds by a triple-alpha reaction $^{4}$He($\alpha\alpha,\gamma)^{12}$C and some of this $^{12}$C converts into $^{16}$O by $^{12}$C($\alpha,\gamma)^{16}$O reaction.
    The helium core burning lasts for $\sim$ $1.3 \times 10^{6}$ yr in the 15 $M_\odot$ model, and a carbon-oxygen core (CO core) is formed.
    After the depletion of helium at the core, the star moves on to a helium-shell burning stage.
    In this convective helium layer, 1-10\% of the helium is burned into carbon.

    \PRF{Stars} of high mass and metallicity experience a strong wind mass-loss in the RSG phase, which strip\PRF{s} their hydrogen envelope and helium layer \cite{Crowther2007}.
    \PRF{These stars} form Wolf-Rayet (WR) star\PRF{s}, which \PRF{have} smaller \PRF{radii} and higher surface temperature\PRF{s} than do RSG\PRF{s}.
    WR stars are also produced through the violent interaction with companions in binary systems.

    In a massive star, the temperature reaches the threshold for carbon burning during the \PRF{contraction} of the CO-core.
    The central temperature becomes $\sim$ 7$\times 10^{8}$ K at the onset of carbon burning, and the carbon burning lasts for $\sim$ 1800 yr in the 15 $M_\odot$ model.
    An ONe core is formed by the production of $^{20}$Ne and $^{24}$Mg via $^{12}$C+$^{12}$C fusion reactions.
    This is followed by several \PRF{phases of} carbon-shell burning for 500 yr, and \PRF{this} carbon-shell burning extend\PRF{s} the ONe core.
    The energy released by neutrinos becomes important at this stage.
    The dominant emission process \PRF{in massive CO cores} is pair annihilation.
    In some lighter stars with higher electron degeneracy, the contribution of plasmon decay becomes larger for neutrino cooling in the central region.
    These neutrinos efficiently cool down the core and suppress \PRF{further} increase of \PRF{the} central temperature.
    \PRF{Carbon} burning then begins at an off-center part of the core, and its burning front gradually moves to the center.
    The 9~$M_\odot$ progenitor in Figure~\ref{ref:rhoc_tc} \PRF{depicts} a lower central temperature at the carbon burning \PRF{phase} compared with the other progenitors. 
    

    Neon ignites when the central temperature reaches $\sim$ 1.4$\times 10^{9}$ K.
    Neon 20 is decomposed into $^{16}$O and $^{4}$He via photo-disintegration and \PRF{is} also converted to $^{24}$Mg and $^{28}$Si through $\alpha$-capture reactions.
    The neon core burning continues for $\sim$ 1 yr in the 15~$M_\odot$ model, and \PRF{the} core is mainly composed of oxygen and silicon (OSi core).

    When the central temperature \PRF{reaches} $1.6 \times 10^{9}$ K, oxygen core burning starts.
    In the 15 $M_\odot$ star model, it takes 9 and 14 months for the contraction\PRF{s} of the OSi core and the oxygen core burning, respectively.
    Oxygen fusion reactions and \PRF{subsequent} reactions produce Si, S, and other intermediate elements.
    A silicon core (Si core) is then formed and grows through \PRF{further} O-shell burnings.

    After the Si core gradually contracts for 2 months, Si-core burning, and an Fe core is formed that consists of iron-group elements.
    The central temperature reaches $\sim$ 3$\times 10^{9}$ K at this time. 
    Silicon \PRF{is} almost depleted \PRF{in} 4.5 days in the 15 $M_\odot$ model, and the Fe core and surrounding Si-rich layer\PRF{s} grow up through Si-shell and O-shell burnings, respectively.
    During contraction of the Fe core, the iron-group elements start to be photo-disintegration, which is an endothermic reaction.
    In addition, neutrinos are emitted by EC reactions in the Fe core, where nuclear statistical equilibrium (NSE) is achieved.
    \textcolor{black}{In the NSE region, the system reaches the minimum of its free-energy and the nuclear composition is unchanged on the same matter condition.}
    These reactions reduce the electron fraction and in turn the \PRF{electron degenerate} pressure decreases.
    These two phenomena enhance core contraction, and the central core begins to collapse 21 hours after silicon core burning.
    At a certain point in the Si-shell burning, EC dominates pair annihilation, which is the main emission process until this time \cite{Kato2017}.


    If the mass of the ONe core is smaller, the evolution at neon burning is different from that of normal massive stars \cite{Nomoto1988,Woosley2015,Suwa2015}.
    The increase \PRF{in} the central temperature is suppressed even during core contraction because of the strong cooling by neutrino emission via plasmon decay.
    The neon ignition hence occurs at an off-center region and causes strong neon-shell burning.
    A convective OSi layer is formed around the ONe core.
    The burning front gradually moves inward, and the neon burning moves to the off-center oxygen burning.
    If the burning front reaches the center before silicon burning \PRF{starts}, \PRF{an} OSi core is formed.
    \PRF{Silicon} burning also occurs at an off-center region, and an Fe-layer is formed on the ONe or Si core.
    The Fe layer \PRF{forms} the Fe core when the burning front reaches the center.
    The central core begins to collapse in the same manner as that \PRF{of} more massive ONe cores.

    When the \PRF{mass of the} ONe core is smaller than the critical mass for neon ignition, 1.35 -- 1.37 $M_\odot$ \cite{Nomoto1984,Schwab2015,Suwa2018}, the maximum temperature of the star does not reach the threshold for neon ignition, and the ONe core continues to contract.
    In this case\PRF{,} the star evolves to a super asymptotic giant branch star.
    During this phase, the ONe core grows through helium-shell and carbon-shell burnings, while the hydrogen envelope is stripped \PRF{away} by a strong mass loss \cite{Siess2010}.
    When the total mass of the star becomes smaller than the critical mass, the star gradually cools down to an ONe white dwarf (WD).
    \textcolor{black}{In contrast, when the ONe core mass of such a star exceeds the critical mass, the star becomes an ECSN.}
    The lowest initial mass for ECSN-progenitors is suggested to be $\sim$ 8 -- 9.5 $M_\odot$, and it depends on the treatment of a convective boundary \cite{Doherty2015}.
    The 9~$M_\odot$ progenitor in Figure~\ref{ref:rhoc_tc} is an example of this evolutionary path.
    The central temperature decreases by neutrino emission via plasmon decays \PRF{and} increases again up to \PRF{$\sim\ 2 \times 10^{9}$ K, which is} the threshold for neon ignition, through ECs of $^{23}$Na, $^{24}$Mg, $^{24}$Na, $^{20}$Ne, and $^{20}$F during core contraction \cite[e.g.,][]{Miyaji1980,Takahashi2013,jones2013}.
    This is accelerated by the \PRF{successive} oxygen and silicon burning, and the temperature exceeds $10^{10}$ K.
    \PRF{Deflagration} starts to propagate outward to convert the ONe core to \PRF{an} Fe core \cite{Nomoto1984,Takahashi2013,Takahashi2019}.
    The Fe core \PRF{then} collapses \PRF{to} an ECSN.
    \textcolor{black}{The explosion of ECSNe is expected to be weaker than that of FeCCSNe.
    However, it is difficult to distinguish it from that of low-mass FeCCSNe.}
    A thermal explosion by the propagation of deflagration wave\PRF{s} is another possibility \PRF{for} ECSNe \cite{Jones2016}.
    In this case\PRF{,} a part of the outer layer could be blown off, with \PRF{an} ONe(Fe) WD left as a remnant.
    This scenario seems to be favorable for star\PRF{s} with low central \PRF{densities} at neon ignition \cite{Leung2019}.

    \subsection{Neutrino emission}

    Pre-SN neutrinos have a key role in \PRF{the} evolutions of massive stars because they are efficiently emitted from the stellar core, \PRF{taking} energy away.
    Neutrino reactions are mainly classified into thermal pair emission and nuclear weak interactions.
    \textcolor{black}{Neutrino luminosities from thermal emission dominate the photon luminosity after carbon burning begins; nuclear weak interactions enhance neutrino emission after heavy element synthesis has commenced.}
    
    \subsubsection{Thermal pair emission}
    There are four processes responsible for neutrino emission:
    \begin{itemize}
            \item{electron-positron pair annihilation (pair)}
            \begin{equation}
                e^+ +e^- \longrightarrow \nu +\bar{\nu} 
            \end{equation}
            \item{plasmon decay (plasmon)}
            \begin{equation}
                \gamma^* \longrightarrow \nu +\bar{\nu}
            \end{equation}
            \item{photo neutrino (photo)}
            \begin{equation}
                e^- + \gamma \longrightarrow e^- + \nu + \bar{\nu}
            \end{equation}
            \item{bremsstrahlung (brems)}
            \begin{equation}
                e^- + (Z,A) \longrightarrow e^- + (Z,A) + \nu + \bar{\nu}
            \end{equation}
        \end{itemize}
    In the above expressions, $Z$ and $A$ are the atomic number and the mass number of nuclei, respectively.
    These processes produce all flavors of neutrinos.
    The reaction rates depend mainly on three hydrodynamical variables: density $\rho$, temperature $T$ and electron fraction $Y_e$ (or chemical potential of electrons).
    Having in mind applications to stellar evolution calculations, Itoh {\it et al.} \cite{itoh1996} obtained useful fitting formulas \PRF{for} the energy loss rates \PRF{in} these processes. 
    They also investigated which reaction is dominant for a given combination of density and temperature.
    According to their results, pair annihilation and plasmon decay \PRF{are} the most important neutrino emission processes \PRF{in} the evolutions of massive stars.
    This finding \PRF{has been} confirmed by other works \cite{odrzywolek2009,Kato2015,Patton2017b,Guo2016}.
    
    The emissivity of pair annihilation is highly sensitive to temperature with the increase in the number of electron-positron pairs.
    The dependence on $\rho Y_e$ is much less drastic: The emissivity decreases with its value because the number of electron-positron pairs is reduced due to the Fermi-degeneracy of electrons.
    The number spectrum of plasmon decay \textcolor{black}{neutrino emission} is much less sensitive to temperature than for pair annihilation, but it depends more on $\rho Y_e$. 
    The peak energy is considerably smaller in plasmon decay. 
    This fact has an important implication for regarding observability of the neutrinos emitted by this process \PRF{in} neutrino detectors.
    
    It should be noted that we do not get neutrino spectra \textcolor{black}{directly} from calculations of stellar evolutions.
    Hence, we evaluate the luminosities and spectra of neutrinos in post \PRF{processing}; that is, we extract \PRF{the} profiles of hydrodynamical variables from stellar evolution calculations.
    Because stars are not homogeneous, we \PRF{first} derive the reaction rates $R(E_\nu, E_{\bar{\nu}},\cos\theta)$, employing local hydrodynamical variables with the energies of neutrino and anti-neutrino $E_\nu$, $E_{\bar{\nu}}$ and the angle between the directions of the neutrino pair $\theta$.
    The numerical formulas \PRF{for pair annihilation \cite{mezzacappa1993,schinder1982} and plasmon decay \cite{Braaten1993}} have been established previously.
    The local spectra for neutrinos and anti-neutrinos are simply given as  integrals of \PRF{the} reaction rates $R$ over the angle and the energy of the partner.
    Finally, the number spectrum $dL_N^\nu/dE_\nu$ is obtained by integration of the local spectra over the star.
    The number luminosity $L_N^\nu$ is a product \PRF{of} the integration of the number spectrum over neutrino energy.

    \subsubsection{Nuclear weak interaction}
    After the silicon burning, nuclear weak interactions can no longer be neglected.
    In particular, once \textcolor{black}{initiated, EC on heavy nuclei} is the dominant reaction.
    It plays an important role in the hydrodynamics of core collapse.
    Although $\beta^+$ decays of heavy nuclei also emit $\nu_e$'s, they are certainly sub-dominant.
    $\bar{\nu}_e$'s are emitted either by positron capture (PC) or $\beta^-$ decay.
        \begin{itemize}
            \item{electron capture (EC)}
            \begin{equation}
                (Z,A) + e^- \longrightarrow (Z-1,A) + \nu_e
            \end{equation}

            \item{$\beta^+$ decay}
            \begin{equation}
                (Z,A) \longrightarrow (Z-1,A) + e^+ + \nu_e
            \end{equation}

            \item{positron capture (PC)}
            \begin{equation}
                (Z,A) + e^+ \longrightarrow (Z+1,A) + \bar{\nu}_e
            \end{equation}

            \item{$\beta^-$ decay}
            \begin{equation}
                (Z,A) \longrightarrow (Z+1,A) + e^- + \bar{\nu}_e.
            \end{equation}
        \end{itemize}
       The reaction rates for these reactions \PRF{have been} calculated in several works [e.g. FFN \citep{fuller1985}, ODA \citep{oda1994}, LMP \citep{langanke2001}, LMSH \citep{langanke2003}].
      The energy spectrum \PRF{for a respective reaction} is reconstructed by the effective $Q$-value method\PRF{, using} the reaction rates and \PRF{the} average energy of neutrinos (See details in References \cite{langanke2001b,sullivan2015,Patton2017b}).

    \subsubsection{Electron capture on free protons}
    Although not abundant, EC on free protons:
        \begin{equation}
            p + e^- \longrightarrow n + \nu_e,
        \end{equation}
    can not be ignored, since the cross section is larger than that of ECs on heavy nuclei \citep{langanke2003}.
    We normally employ the reaction rate \PRF{calculated} by Bruenn {\it et al.} \cite{bruenn1985} \PRF{in} this process.

 　 \begin{figure}
    \centering
    \includegraphics[scale=0.7]{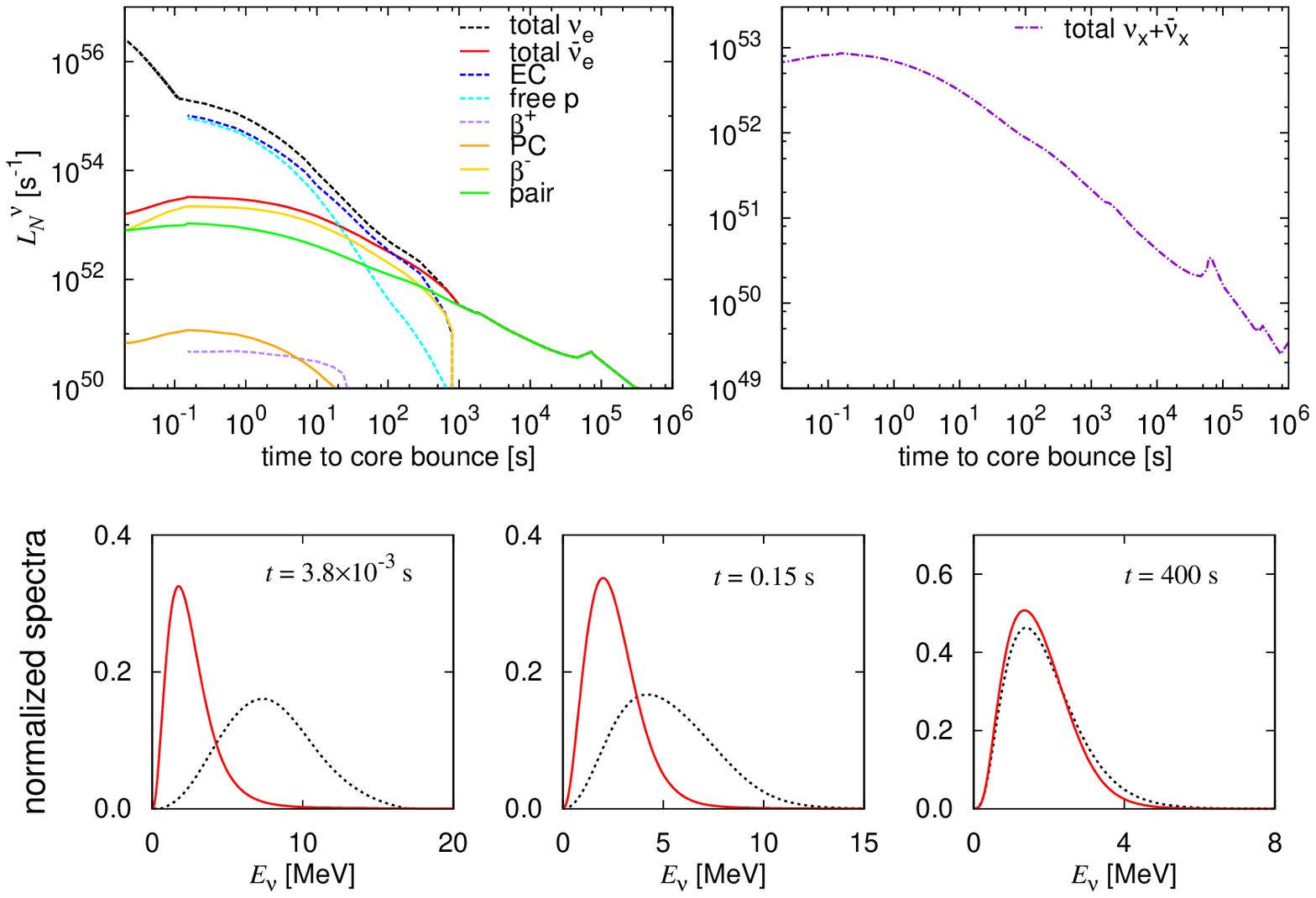}
    \caption{Top: Time evolution of number luminosities for the emission processes of $\bar{\nu}_e$'s (solid) and $\nu_e$'s (dotted) \textcolor{black}{in the left panel and for that of $\nu_x$'s and $\bar{\nu}_x$'s in the right panel for the Kato model \cite{Kato2017}. All luminosities of heavy-lepton (anti-)neutrinos are summed up.}
    Time is measured from core bounce and different colors denote different emission processes. We switch \PRF{from} the quasi-static calculation of stellar evolution to the hydrodynamical simulation for the core-collapse phase at $t\sim10^{-1}$ s. For $\nu_e$'s in the collapse phase, only the total luminosity is shown because it is all that the transport calculations produce. Bottom: Snapshots of \PRF{the} number spectra for $\bar{\nu}_e$'s (red, solid) and $\nu_e$'s (black, dotted) at $t=3.8\times10^{-3},\ 0.15$ and 400 s.
    They are normalized by the number luminosities $L_N^\nu$ at each time point. Neutrino oscillations are not included in all panels of this figure.}
    \label{ref:luminosity}
    \end{figure}
    
\subsection{Luminosities and spectra of pre-SN neutrinos}\label{subsec:L_spec_preSNnu}
  In this section, on the basis of a model from Kato {\it et al.} \cite{Kato2017} we introduce typical properties of pre-SN neutrinos.
  This model considers neutrino emission during both the quasi-static evolutions of progenitors and the hydrodynamical core collapse until $\rho_c=10^{13}\ \rm{g~cm^{-3}}$.
  For the former, they use the stellar evolution model with 15 $M_\odot$, whereas for the latter they conduct a one-dimensional simulation under spherical symmetry to solve radiation-hydrodynamic equations \cite{nagakura2017}.
  It is important to take into account neutrino transport in the core properly, once the density becomes high enough to trap neutrinos. 
  The two evolutionary phases are \PRF{mutually} connected when the central density becomes $\rho_c=10^{10.3}\ \rm{g~cm^{-3}}$, which corresponds to $t\approx 0.1$ s before core bounce.
  The \textcolor{black}{top-left panel} of Figure~\ref{ref:luminosity} shows the time evolution of number luminosities.
  The number luminosity of $\nu_e$'s (black, dotted) gradually increases up to $\sim10^{55}$ and $10^{57}\ \rm{s}^{-1}$ just before core collapse and core bounce, respectively.
  EC on heavy nuclei (blue, dotted) and free protons (cyan, dotted) makes large contributions to \PRF{the total luminosity}.
  The $\nu_e$ emission occurs predominantly in the collapse phase ($t < 0.1$ s), although this phase is much shorter than the progenitor phase that precedes it.
  For $\bar{\nu}_e$'s, the number luminosity (red, solid) reaches a maximum $L_N^\nu\sim10^{53}\ \rm{s}^{-1}$ at the beginning of the core collapse.
  Neutrino emission via pair annihilation (green, solid) dominates the other reactions at $t>400$ s, whereas after this, $\beta^-$ decay (yellow, solid) overcomes \PRF{it}.
  Because of the Fermi-blocking of electrons in the final state, $\beta^-$ decay is suppressed at high densities, at which electrons are strongly degenerate, and the number luminosity of $\bar{\nu}_e$'s \PRF{decreases, eventually leading to} core bounce.
  As a consequence, the progenitor phase is dominant over the collapse phase in $\bar{\nu}_e$ emission.
  \textcolor{black}{The total number luminosity of $\nu_x$'s is shown in the top-right panel of Figure~\ref{ref:luminosity}.
  Because $\nu_x$'s are only emitted via pair annihilation in this phase, the luminosity is much smaller than those of $\nu_e$'s and $\bar{\nu}_e$'s. 
  It is difficult for currently operational as well as planned detectors to detect $\nu_x$'s. They are, however, converted to $\nu_e$'s or $\bar{\nu}_e$'s through neutrino oscillation, and they affect the fluxes of other neutrino flavors on the Earth (see Section~\ref{subsec:sensitivity}).}
  
  Time snapshots of \PRF{the} neutrino spectra at $t = 3.8\times10^{-3},\ 0.15$ and 400 s are shown in the bottom panels of Figure~\ref{ref:luminosity}.
  At $t=$ 400 s, the dominant reaction for $\bar{\nu}_e$'s \PRF{changes}, while the other two time points correspond \PRF{respectively} to the times just before core collapse and core bounce.
  The average energy of $\nu_e$'s reaches $\sim$ 8 MeV before core bounce, whereas that of $\bar{\nu}_e$'s \PRF{remains} $\sim$ 3 MeV.
  The difference in neutrino flavors is caused by \PRF{the difference} of the degeneracy between electrons and positrons.
  \textcolor{black}{The results of the Kato model are available at \citep{katodata}, and readers may refer to these data for the time evolution of neutrino spectra. }
  
  \begin{figure}
 \centering
 \includegraphics[scale=0.5]{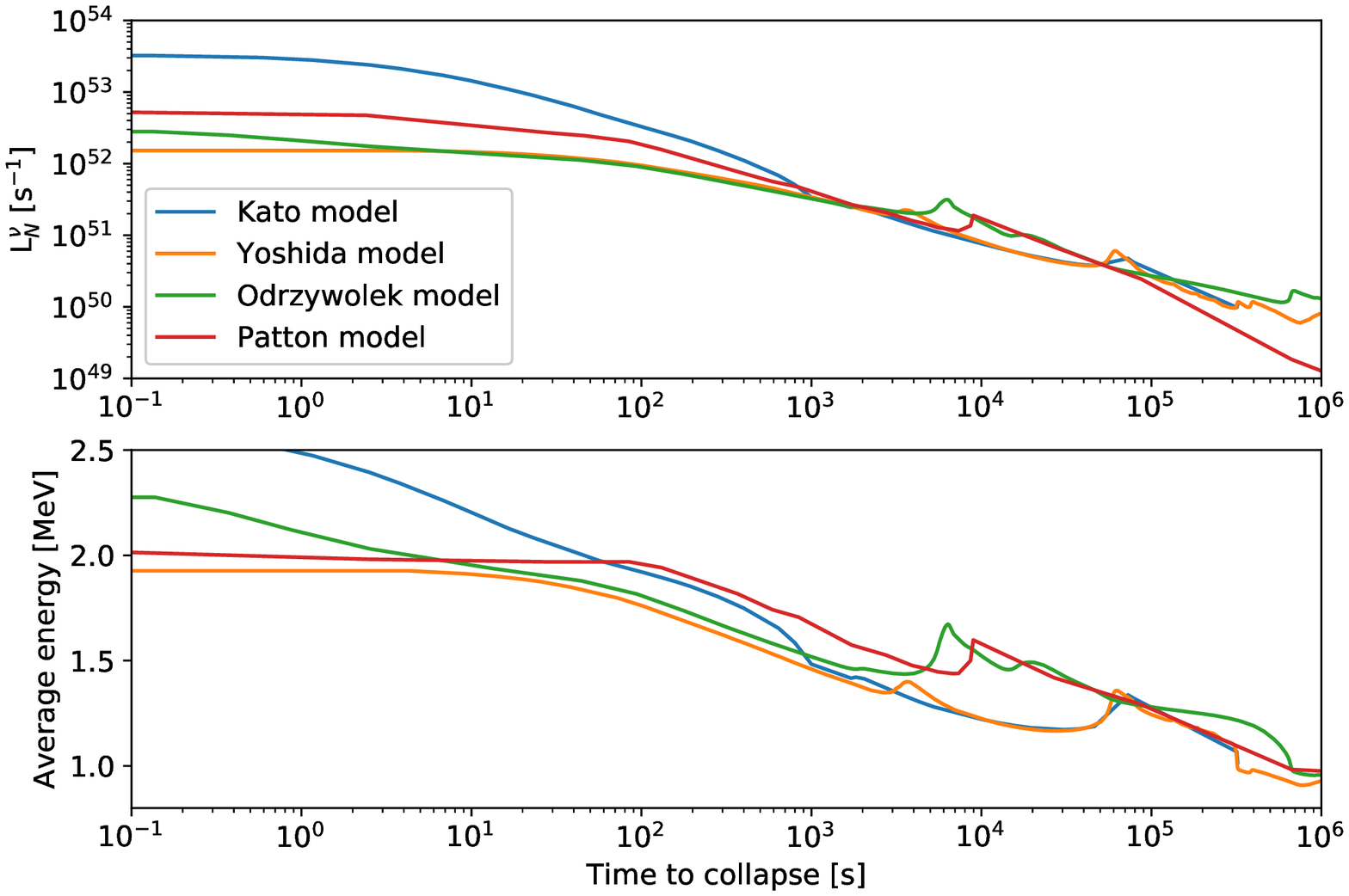}
 \caption{Time evolution of number luminosities (top) and average energies (bottom) of $\bar{\nu}_e$'s for four pre-SN neutrino models \citep{Yoshida2016,Kato2017,Patton2017,odrzywolek2010}. Neutrino oscillations are not included.}
 \label{ref:timeEvolution1}
\end{figure}

  \PRF{The} properties of pre-SN neutrinos depend on stellar models and neutrino processes employed in calculations.
  In this review, we focus on four representative models for pre-SN neutrinos, \PRF{the} Kato \cite{Kato2017}, Yoshida \cite{Yoshida2016}\footnote{https://doi.org/10.5281/zenodo.3778014}, Odrzywolek \cite{odrzywolek2010}\footnote{http://th.if.uj.edu.pl/~odrzywolek/psns/index.html} and Patton \cite{Patton2017} models\footnote{https://doi.org/10.5281/zenodo.2598709}. 
  The four models employ different stellar models; Takahashi models \cite{Yoshida2016,Kato2017} calculated using HOSHI code \cite{Takahashi2016,Takahashi2018,Takahashi2019,Yoshida2019} for the first two and Woosley \cite{Woosley2002} and MESA models \cite{farmer2016} for the latter two.
  The Odrzywolek and Yoshida models take only pair annihilation into their calculations, whereas the Kato and Patton models also include nuclear weak interactions.
  Figure~\ref{ref:timeEvolution1} shows the time evolutions of \PRF{the} number luminosities (top) and average energies of $\bar{\nu}_e$'s (bottom) for the different stellar models with 15 $M_\odot$.
  Here, we focus on the evolution until core collapse.
  We find that the number luminosities of the Kato and Patton models are larger than those of the other two because of nuclear weak interactions.
  Even though both the Kato and Patton models take the same nuclear weak interactions into account, their luminosities and average energies have large differences at $t<100$ s.
  \PRF{This} seems to be due to the different treatments \PRF{of the} mass fractions of nuclei.
  In the Patton model, mass fractions derived from solving nuclear networks with 204 isotopes in both NSE and non-NSE regions \PRF{have been adopted}, whereas the Kato model adopts NSE compositions recalculated for 3928 nuclei with Furusawa equation of state \cite{Furusawa2013} in post \PRF{processing}. 
  The Odrzywolek and Yoshida models, in which only pair annihilation \PRF{is included}, have similar number luminosities except for \PRF{at} $t < 5$ s.
  The deviations in number luminosities and average energies between \PRF{the} two models at $t<5$ s \PRF{seemingly} come from the difference in progenitor models, but \PRF{detailed investigation is necessary to make conclusive remarks.}
  We also find two peaks in the number luminosities at $t\sim 5\times10^{3}$ and $5\times10^4$ s, which correspond to the Si-shell and O-shell burnings, respectively (See Section~\ref{convection}).
  \PRF{Features such as amplitudes and widths of peaks} depend on \PRF{the} pre-SN neutrino models.
  Because these peaks will be confirmed by pre-SN neutrino observations \cite{Yoshida2016}, \PRF{a detailed} comparison of shell burnings among these models is required.
  It should be noted that the origin of the time may be changed due to the theoretical uncertainties in the definition of core collapse.

  In any case, pre-SN neutrinos \PRF{are smaller in number} and \PRF{have} lower energies \PRF{compared with} supernova neutrinos, for which $L_N^\nu \sim 10^{58}\ \rm{s}^{-1}$ and $\langle E_\nu \rangle $ is \PRF{in the order of} tens MeV \citep{wallace2015}.
  \PRF{Therefore,} we need detectors with high sensitivity to low-energy neutrinos and with a low background for the detection of pre-SN neutrinos.
  
%

\subsection{Neutrino oscillation}
Neutrinos have small but finite masses \cite[e.g.,][]{Smy2003, Ashie2005, Abe2014} and can convert their flavors during propagation when mass eigenstates and flavor eigenstates do not match.
This phenomenon is called a neutrino oscillation and affects neutrino fluxes.
Its behavior depends on the surrounding condition.
Two types of oscillation---vacuum oscillation and Mikheyev-Sumirnov-Wolfenstein (MSW) effect via electron forward-scattering \cite{wolfenstein1987}---should be taken into account in the pre-bounce phase.
\textcolor{black}{In the MSW effects, the effective mixing angles depend on the electron number density and reach their maximum at resonance points.
They are located near the boundaries C-He and He-H layers.
Neutrinos are mainly produced inside the resonance points and certainly pass through them.
The mass eigenstate of $\bar{\nu}_e$'s is $\bar{\nu}_{m1}$ and $\bar{\nu}_{m3}$ at birth in the normal and inverted mass orderings, respectively, where $\bar{\nu}_{mi}$ is the $i$-th mass eigenstate of anti-neutrinos.
In the pre-bounce phase, the density scale height at the resonance points is longer than the oscillation length.
The adiabatic approximation is therefore available, and the transition of mass eigenstates never occurs, irrespective of the progenitor types \cite{Kato2015}.
}
Note that collective oscillation \cite[e.g.,][]{duan2010} induced by neutrino self-interaction is negligible here because neutrinos can freely escape from a star, and the self-interaction hardly occurs.

If we take two oscillations into account, the flux of pre-SN $\bar{\nu}_e$'s on the Earth is shown as \cite{Dighe:1999bi},
\begin{equation}
   F_{\bar{\nu}_e}(E_\nu,t,d) =  p F^0_{\bar{\nu}_e}(E_\nu,t,d) + (1-p)F^0_{\bar{\nu}_x}(E_\nu,t,d), \label{eq:nuflux}
\end{equation}
where $F^0_{\bar{\nu}_\alpha}(E_{\nu},t,d)$ is the flux of $\alpha$ flavor without neutrino oscillations,
$p = \cos^2\theta_{12}\cos^2\theta_{13} ~(\sin^2\theta_{13})$ is the survival probability for the normal (inverted) mass ordering in the adiabatic limit,
$E_{\nu}$ is the neutrino energy, $t$ is the time to collapse, $d$ is the distance to the star, $\theta_{12}$ and $\theta_{13}$ are mixing angles.
We employ this flux in estimating the number of IBD events in Section~\ref{subsec:sensitivity}.



\section{PRE-SUPERNOVA NEUTRINO DETECTION AND SN ALARM} \label{sec3}


In this section, we introduce neutrino detectors \textcolor{black}{that could} detect pre-SN neutrinos, and a scheme to estimate their detection ranges of distance and time.
We then \textcolor{black}{apply} this scheme to \textcolor{black}{the} four pre-SN neutrino models \textcolor{black}{with 15 $M_\odot$} introduced in the previous section.
\textcolor{black}{We also} discuss the current status and 
\textcolor{black}{some} future prospects of \textcolor{black}{an} SN alarm with pre-SN neutrinos.

\subsection{Detectors and pre-SN neutrino observations}
We first introduce the current and future detectors, mainly focusing on the detection of pre-SN neutrinos.

\subsubsection{KamLAND} 
KamLAND is a 1-kt neutrino detector located at the Kamioka mine in Japan \citep{suzuki2014}. 
KamLAND can detect \textcolor{black}{pre-SN} $\bar{\nu}_e$ events via IBD in low-background \textcolor{black}{conditions, owing} to the selection of a DC pair of prompt and delayed events with \textcolor{black}{temporal} and spatial correlations. 
The \textcolor{black}{prompt event indicates scintillation light from} the kinetic energy of positron and two 511-keV $\gamma$-rays via the pair annihilation of a positron and an electron. 
\textcolor{black}{The delayed event is} a 2.2-MeV $\gamma$-ray through neutron capture on a proton. 
This \textcolor{black}{DC method} can reject most accidental backgrounds.

Unavoidable backgrounds for pre-SN neutrinos are \textcolor{black}{the terrestrial} neutrinos emitted \textcolor{black}{in the} decays of nuclei in reactors and \textcolor{black}{radionuclides throughout the Earth's interior}. 
The largest uncertainty is \textcolor{black}{due to the statuses of the reactors in Japan.} 
\textcolor{black}{The Great East Japan Earthquake in 2011 lowered the the reactor background. The background rate is 0.07 events/day}, with a fiducial volume of $\sim$ 0.7~kt in the energy range for pre-SN neutrinos, $0.9 \leq E_p \leq 3.5$~MeV \citep{Asakura2016}. 
\textcolor{black}{Here,} $E_p$ is the visible energy of the prompt signal. 
\textcolor{black}{Before the earthquake,} 
the background rate \textcolor{black}{was} 0.35 events/day. 

The KamLAND collaboration evaluated the detection possibility of pre-SN neutrinos from nearby massive stars and found that KamLAND will detect pre-SN neutrinos from stars within 690~pc with the 3$\sigma$ significance \citep{Asakura2016}\textcolor{black}{, assuming} low-reactor backgrounds, based on the 25~$M_\odot$ model developed by Odrzywolek \& Heger~\cite{odrzywolek2010}. 

\subsubsection{SNO+}
Sudbury Neutrino Observatory plus~(SNO+) is an experiment intended to search for neutrino-less double-$\beta$ decay of $^{130}$Te, using the underground equipment already installed for SNO at SNOLAB in Sudbury, Canada \cite{Andringa:2015tza}. 
It is designed to use 780~t liquid scintillator.  
\textcolor{black}{SNO+ will use IBD to detect pre-SN $\bar{\nu}_e$'s.}  
Without $^{130}$Te, the expected total number of reactor and geo-neutrino events with IBD \textcolor{black}{is expected to be} $\sim$ 88.5
Terrestrial Neutrino Units (TNU), which corresponds one DC event per $10^{32}$ free protons per year~\cite{Baldoncini2016}, 
in a low energy region. 
Roughly speaking, the DC event rate \textcolor{black}{from reactor and geo-neutrinos} is about 0.16 events/day. 
\textcolor{black}{This is the background rate for pre-SN neutrinos.}
Without Te, the sensitivity of SNO+ is similar to that of KamLAND. 
\textcolor{black}{However, a}fter loading Te, the sensitivity will become worse in SNO+ because of the increase \textcolor{black}{in} accidental backgrounds through two-neutrino double-$\beta$ decays of $^{130}$Te. 

\subsubsection{JUNO}
JUNO is a \textcolor{black}{20-kt} multi-purpose underground liquid scintillator detector \citep{an2016}. 
It is designed to deploy a single (far) detector at baselines of $\sim$ 53 km from both the Yangjiang and the Taishan reactors to confirm neutrino mass ordering and precisely measure oscillation parameters with reactor neutrinos.
Pre-SN neutrinos can be detected via IBD at JUNO. 
\textcolor{black}{The cumulative} expected numbers of DC events \textcolor{black}{on the last day before core collapse at $d = 1$~kpc} are 6.1 (1.9), 12.0 (3.6), 20.5 (5.9) and 24.5 (7.0) for the normal (inverted) mass ordering using stellar models \cite{Woosley2002} with 12, 15, 20 and 25~$M_\odot$, respectively \cite{Guo2019}. 
This estimation assumes a detection efficiency of 0.83 and a $\bar{\nu}_e$ energy window of 1.8 $\leq E_{\bar{\nu}_e} \leq$ 4 MeV.  
\textcolor{black}{The corresponding} reactor and geo-$\bar{\nu}_e$ events are 15.7 and 1.1, respectively. 
If \textcolor{black}{the distance to the source is known exactly,} JUNO can determine the neutrino mass ordering at a $>$ 95\% confidence level for the star\textcolor{black}{s} located within 440--880~pc. 
It is, however, too optimistic \textcolor{black}{to assume that this situation is realized,} and other model-independent method is discussed with a combination of IBD and Electron Scattering~(ES) events \cite{Guo2019}.  
If the cosmogenic backgrounds for the ES events will be reduced by a factor of 2.5--10,  
a model-independent determination of the neutrino mass ordering is possible with pre-SN neutrinos from Betelgeuse. 

\subsubsection{SK and SK-Gd}
SK is a water Cherenkov detector located in the Kamioka mine in Japan, and IBD is available to detect $\bar{\nu}_e$'s \citep{abe2014b}. 
Unfortunately, \textcolor{black}{the detection efficiency of} delayed 2.2-MeV $\gamma$-rays through neutron capture on protons is 
\textcolor{black}{only 17.74\% for a  supernova relic neutrino search~\cite{Zhang:2013tua}. } 
In addition, only a small part of a prompt event exceeds \textcolor{black}{the detection threshold of 3.5~MeV} 
in the electron kinetic energy and $\sim$ 5.3~MeV in the neutrino energy. 
This is much larger than the threshold for IBD (\textcolor{black}{$E_{\bar{\nu}_e} =  1.8$~MeV} ).
\textcolor{black}{Moreover, prompt} events \textcolor{black}{generated from pre-SN $\bar{\nu}_e$'s} are,  buried in backgrounds without the delayed events.
\textcolor{black}{Therefore,} SK is not currently suitable for the detection of pre-SN neutrinos. 

Recently, SK-Gd was approved.
Neutrons produced through IBD are captured by Gd and 8-MeV $\gamma$-rays are released. 
These delayed events will be visible. 
\textcolor{black}{However, the} detection efficiency of prompt signals  is still low, and the SK collaboration proposed two analyses for pre-SN neutrino events \cite{simpson2019}. 
One is the channel with DC events\textcolor{black}{, even if only a small number} of prompt events is visible. 
A typical selection efficiency of DC events is 3.9--6.7\%, with a background rate of 24--56 events/day. 
The other is the analysis only for the delayed neutron events. 
\textcolor{black}{The selection efficiency is estimated to be} 7.3--10\%, with a background rate of 132--280 events/day. 
Based on \textcolor{black}{this} analysis, \textcolor{black}{the} maximum detection range is $\sim$ 600~pc with \textcolor{black}{one false alarm per yr}, assuming the normal mass ordering based on the 15 and 25~$M_\odot$ models in Patton {\it et al.} \cite{Patton2017}. 
\textcolor{black}{The first phase of SK-Gd will start with 10 tons of Gd${_2}$(SO$_4$)$_3$ (0.01\% Gd concentration) in 2020 (one-tenth of the final goal).} 

\subsubsection{Dark matter detectors}
Large dark matter experiments \textcolor{black}{may potentially detect} pre-SN neutrinos because of the combination \textcolor{black}{of} 
very low detection thresholds (around or below the keV level) and \textcolor{black}{CE$\nu$NS} \cite{Raj2019}. 
One advantage in these experiments is a sensitivity to all six flavors of neutrinos\textcolor{black}{, from which} complementary observations of pre-SN neutrinos are expected. 
For example, a proposed argon detector with a target mass of 300~t and an energy 
threshold of 0.6~keV would detect 23.6 (68.4) events within a 12-hour time window before collapse of a 15~$M_\odot$~(30~$M_\odot$) 
\textcolor{black}{star at a distance within} $d=$ 200~pc.
In such \textcolor{black}{a} configuration, \textcolor{black}{the} detection significance exceeds 3$\sigma$ $\mathcal{O}(10)$ hours prior to collapse in the 30~$M_\odot$ star, whereas the same situation is satisfied within an hour before collapse for the 15~$M_\odot$ star. 

\subsubsection{DUNE}
DUNE \textcolor{black}{consists} of four 
\textcolor{black}{10-kt} liquid argon time projection chambers, \textcolor{black}{located} at SURF in South Dakota \cite{acciarri16}. 
\textcolor{black}{It is} designed as a far detector for the Long-Baseline Neutrino Facility (LBNF). 
The primary \textcolor{black}{objectives of} DUNE are to determine neutrino mass ordering, to measure a CP violation in the neutrino sector, \textcolor{black}{and to measure supernova neutrino bursts after core bounce.} 
Even so, DUNE will play an important role in pre-SN neutrino observations because it has \textcolor{black}{the} highest \textcolor{black}{sensitivity} to $\nu_e$'s among \textcolor{black}{all the currently planned} neutrino detectors. 
In particular, DUNE will detect \textcolor{black}{tens of} thousands of charged current (CC) events \textcolor{black}{before core bounce for the ECSN-progenitor} with $M_{\rm{ZAMS}} = 9~M_\odot$,
assuming that the energy threshold is $E_\nu = 10.8$~MeV \cite{Kato2017}.

\subsection{Detector sensitivity}
\label{subsec:sensitivity}

\textcolor{black}{The sensitivity of detectors to pre-SN neutrinos has been discussed in the literature}~\cite{simpson2019,Kato2015,Yoshida2016,Kato2017,Raj2019,Asakura2016,Guo2019}. 
However, different studies have often used different pre-SN models, detection statistics, and/or time windows. 
Here, we present a unified method and discuss detector sensitivity with IBD on SK-Gd, KamLAND, and JUNO. 
\textcolor{black}{In this study, we estimate the pre-SN $\bar{\nu}_e$ flux using eq.(\ref{eq:nuflux}) with $\cos^2\theta_{12} = 0.692$ and $\cos^2\theta_{13} = 0.9766$ \cite{Agashe:2014kda}.
To be technically accurate, we assume that $t=0$ is the time, at which these studies stop the calculations of stellar evolutions.
We regard this time as the onset of collapse.
We also assume that the detector background rate is constant.
Although we skip the sensitivity of DUNE and large dark matter detectors because of a lack of detailed information, 
DUNE has a unique advantage for pre-SN neutrino detection because it is sensitive to $\nu_e$'s via CC interactions, while large dark matter detectors can detect all flavors of neutrinos. }

\begin{figure}
 \centering
 \includegraphics[scale=0.5]{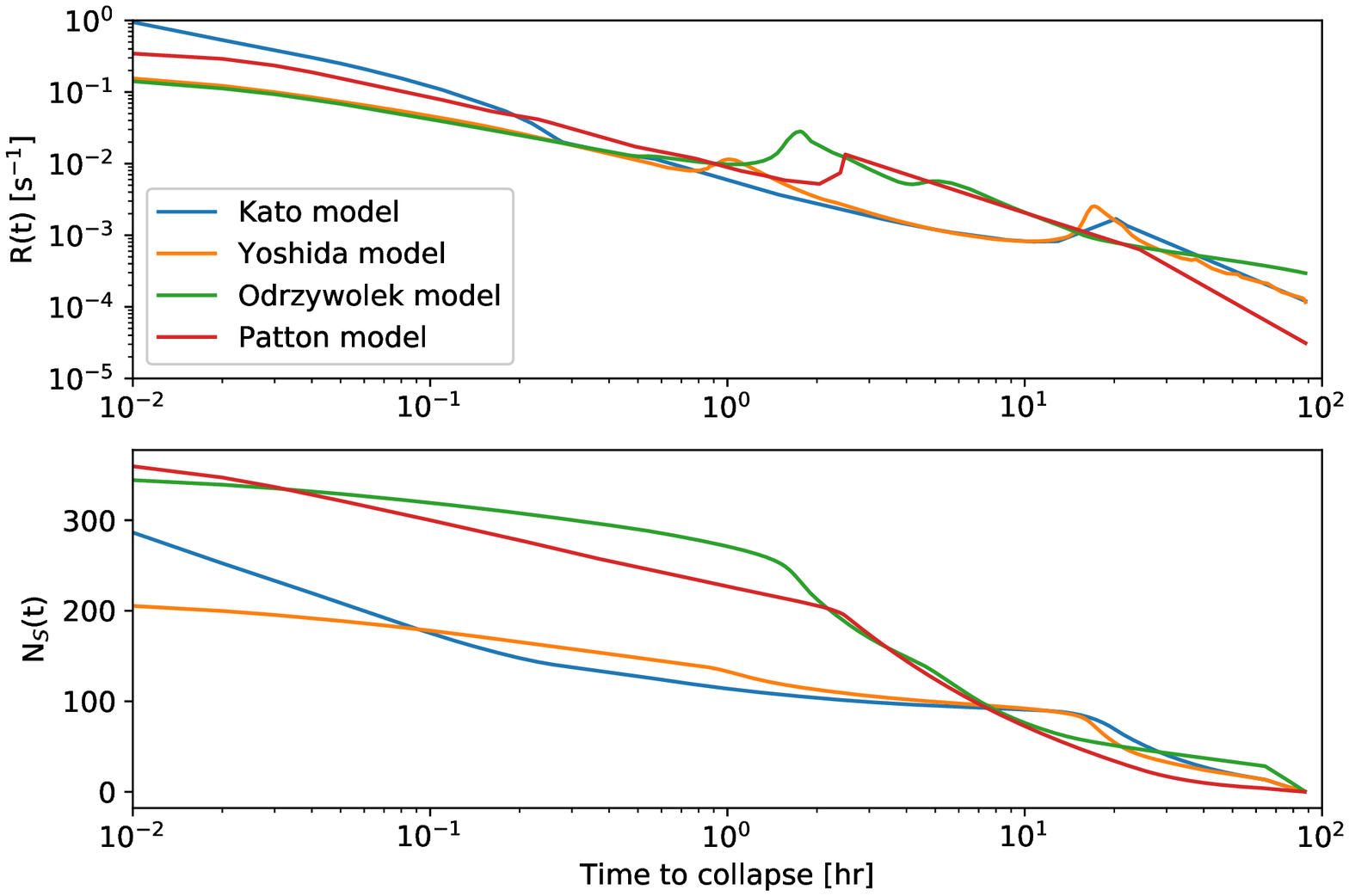}
 \caption{Time evolutions of event rates (top) and 
 cumulative numbers of DC events (bottom) at JUNO.
 We adopt four pre-SN neutrino models \citep{Yoshida2016,Kato2017,Patton2017,odrzywolek2010} to the estimation and assume that neutrinos have normal mass ordering. The time window is 24~hr, and the distance is $200$~pc. }
 \label{ref:timeEvolution2}
\end{figure}

An event rate at a time $t$ is written as
\begin{equation}
    R(t,d) = N_P \int F_{\bar{\nu}_e}(E_\nu,t,d)\sigma(E_\nu) \epsilon(E_\nu) dE_\nu ,
\end{equation}
where $N_P$ is \textcolor{black}{the} number of target protons, $\sigma(E_\nu)$ is \textcolor{black}{the} cross section of IBD and 
$\epsilon(E_\nu)$ is \textcolor{black}{the} detection efficiency as a function of \textcolor{black}{the} neutrino energy $E_\nu$. 
The time evolutions of $R(t,d)$ for the four pre-SN models at JUNO with $d=200$~pc are shown in the top panel of  Figure~\ref{ref:timeEvolution2}.
The time profiles exactly show those of the number luminosities and average energies (i.e., the peaks associated with shell burnings) in Figure~\ref{ref:timeEvolution1}.
We find that $R(t,d)$ is generally smaller than one and hence we have to discuss the pre-SN neutrino detection with a cumulative number of events defined as 
$\langle N_s(t,d) \rangle = \int^{t+t_w}_{t}  R(t^\prime,d)dt^\prime$, with a fixed time window of $t_w$. 
Here $\langle \cdot \cdot \cdot  \rangle$ indicates the ensemble average with repeated measurements.
We show the time evolution of $\langle N_s \rangle$ with $t_w=24$ hr and $d=200$~pc for the normal ordering in the bottom panel of Figure~\ref{ref:timeEvolution2}.
In the Patton and Odrzywolek models, we find the steep gradient in $\langle N_s \rangle$ due to the shell burnings at $t\sim2-3$ hr.
In the Kato and Yoshida models, the $\langle N_s \rangle$'s have their extrema at $t\sim20$ hr because they decrease after the peaks associated with shell burnings are out of the range for $t_w$.


\begin{table}[htb]
\caption{Summary of detector parameters.
Detector size is defined as the full volume for SK-Gd and JUNO and as the fiducial volume for KamLAND.}
  \begin{tabular}{lcccc} \label{tab:detector}
    Detector name & Detector size & BG [event/day] & Efficiency & Estimation \\
    \hline
    SK-Gd (neutron) & 32.5~kt & 132 & 10\%  & Optimistic \\
    SK-Gd (neutron) & 32.5~kt & 280 & 7.3\%  & Pessimistic \\
    SK-Gd (DC) & 32.5~kt & 24 & Black curve in Fig.~8 of Ref.~\cite{simpson2019}  & Optimistic \\
    SK-Gd (DC) & 32.5~kt & 56 & Blue curve in Fig.~8 of Ref.~\cite{simpson2019}  & Pessimistic \\
    KamLAND & 0.7~kt & 0.07 & Fig.~4 of Ref.~\cite{Asakura2016} & Optimistic \\
    KamLAND & 0.7~kt & 0.35 & Fig.~4 of Ref.~\cite{Asakura2016} & Pessimistic \\
    JUNO & 20~kt & 16.8 & 0.73 
  \end{tabular}
\end{table}

\textcolor{black}{The} cumulative number of events $N(t)$ actually observed at neutrino detectors, follows the Poisson probability $P(N(t), N_{\textrm{ex}}(t,d))$, with an average $N_\textrm{ex}(t,d) = \langle N_\textrm{s}(t,d) \rangle  + \langle N_b \rangle$, 
assuming that $N(t_1)$ and $N(t_2)$~($t_1 \neq t_2$) are independent.  
\textcolor{black}{Here,} $\langle N_b \rangle$ is \textcolor{black}{the} estimated number of background events with a time window of $t_w$.
If we assume \textcolor{black}{a} 50\% detection efficiency, \textcolor{black}{the} cumulative number of events $N_{50}(t,d)$ is defined with the given $N_\textrm{ex}(t)$ \textcolor{black}{as} 
\begin{equation}
    \int_0^{N_{50}(t,d)} P(N(t), N_{\textrm{ex}}(t,d)) dN = 0.5. 
\end{equation}
We then get an accidental detection probability $A(t,d)$ for $N_{50}(t,d)$, 
\begin{equation}
    A(t,d) = \int^\infty_{N_{50}(t,d)} P(N, \langle N_{b} \rangle) dN. 
\end{equation}
Generally, a false alarm rate per year, $F(t,d)=A(t,d) \times 365 \times 24/t_w$, is preferable for the discussion of the accidental detection to $A(t,d)$ itself. 

Detector parameters used in this section are summarized in Table~\ref{tab:detector}. 
We define detector size \textcolor{black}{as} the full volume for SK-Gd and JUNO and \textcolor{black}{as} the fiducial volume for KamLAND.
\textcolor{black}{We make optimistic~(high detection efficiency and low backgrounds) and pessimistic (low detection efficiency and high backgrounds) estimations for SK-Gd from Reference \cite{simpson2019}.}
We also employ optimistic (low-reactor) and pessimistic (normal-reactor) background conditions for KamLAND.
We consider only reactor and geo-neutrinos as the background for JUNO, although this assumption is less realistic than that \textcolor{black}{for} SK-Gd and KamLAND.

\begin{figure}
 \centering

 \includegraphics[scale=0.4]{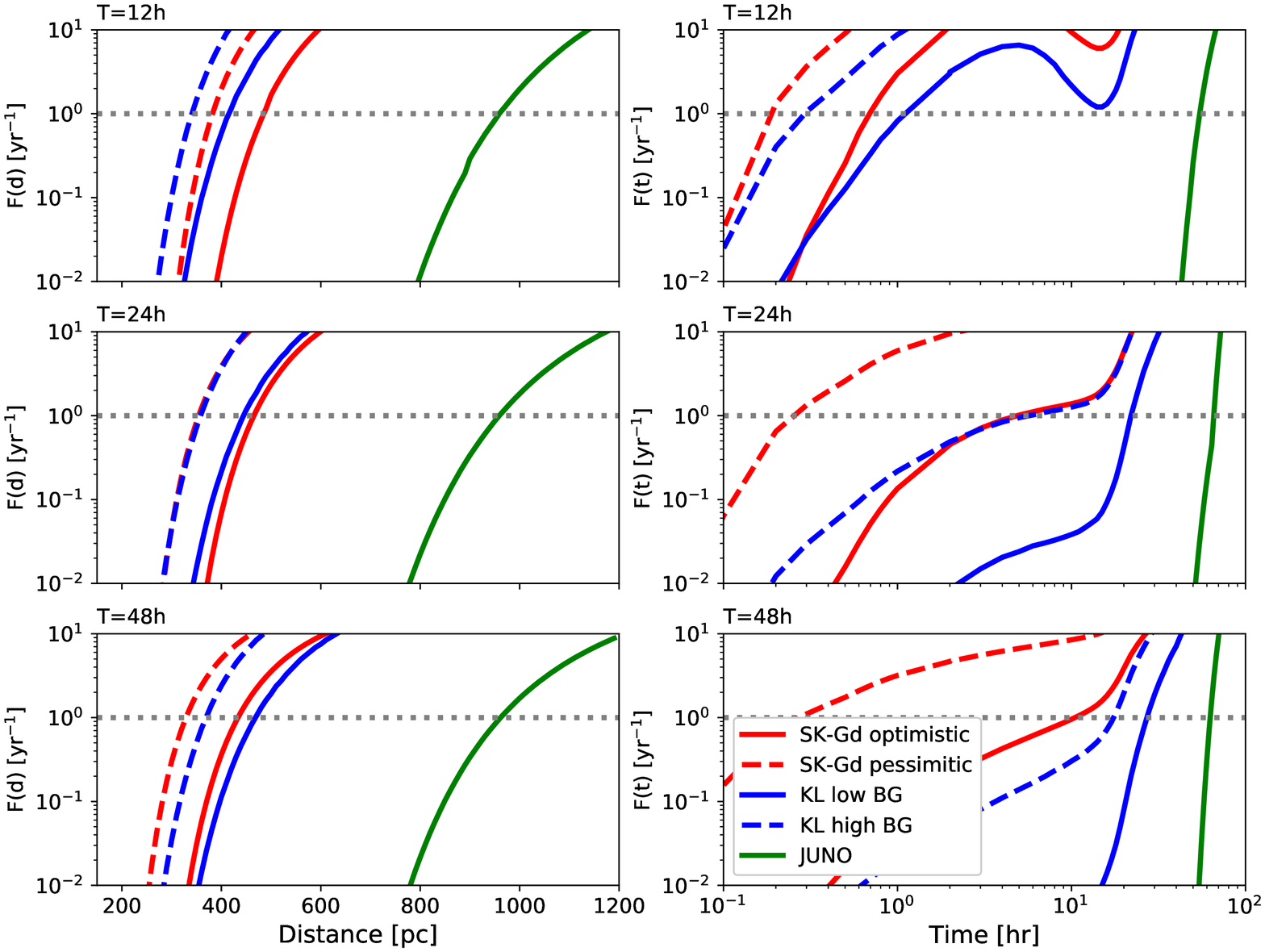}
  \caption{False alarm rate as a function of distance (left) and time (right) with the Kato model \citep{Kato2017} for normal mass ordering.
  It is assumed that the supernova occurs at 200 pc in the right panels.
  The time windows are defined as $t_w$=12, 24 and 48~hr from top to bottom. 
  The dotted lines show a false alarm rate with 1~yr$^{-1}$. }
 \label{ref:detectionRange}
\end{figure}

The left panels of Figure~\ref{ref:detectionRange} show $F(d)$ at $t=0.01~{\textrm{hr}}$ for the Kato model as a function of distance in SK-Gd, KamLAND, and JUNO 
with $t_w=12$, 24 and 48~hr.
We assume that neutrinos have the normal ordering.
Here, we focus on the time range $t>0.01$~hr because the exact definition of $t=0$ is uncertain and depends on models as we mentioned.
For SK-Gd, we assume $F_{\rm{SK}}(t,d) = A_{\rm{n}}(t,d) A_{\rm{DC}}(t,d)\times 365/t_w$, 
where $A_{\rm{n}}(t,d)$ and $A_{\rm{DC}}(t,d)$ are 
\textcolor{black}{the} accidental detection probabilities for only neutron events and DC events, respectively.
If we require $F(t,d)$ = 1 yr$^{-1}$ for a detection, SK-Gd and KamLAND have a similar sensitivity to pre-SN neutrinos and have enough capability to predict the supernova explosion of Betelgeuse (200~pc) before collapse. 
JUNO has the highest sensitivity, and the distance can be extended to $\sim$ 1~kpc. 
Regarding the capabilities of an SN alarm with pre-SN neutrinos, the right panels of Figure~\ref{ref:detectionRange} show $F(t)$ as a function of time to collapse, 
assuming that the Betelgeuse supernova occurs at $d=200$~pc. 
The earliest alarm will be \textcolor{black}{triggerd} by the detection at JUNO $\sim$ 78~hr prior to collapse, when $F(t,d)$ = 1 yr$^{-1}$.
KamLAND is preferable for triggering an early alarm to SK-Gd for the Kato model.

Previous studies \textcolor{black}{have employed} windows of $t_w=12$ and 48~hr in estimating false alarm rates for SK-Gd and KamLAND, respectively.
An optimization of $t_w$ is, however, unclear.
Even focusing on SK-Gd, the longest detection distance is derived when $t_w=12$~hr, 
whereas the best $t_w$ for the alarm time 
depends on the models \textcolor{black}{considered}. 
The most promising $t_w$, moreover, depends on detectors.
Actually, the detection distance becomes  \textcolor{black}{the} longest in the case of $t_w=48$ hr at KamLAND.
It is obvious that $\langle N_s \rangle$ and $\langle N_g \rangle$
increase with a time window $t_w$, and the most promising $t_w$ is determined by keeping a balance between their increments with $t_w$.
At KamLAND, the low $\langle N_g \rangle$ is achieved even for $t_w=48$ hr, and it provides us the best sensitivity.
The shorter \textcolor{black}{the} time window is, \textcolor{black}{the} more promising at SK-Gd because of the higher $\langle N_g \rangle$.
JUNO has intermediate properties between SK-Gd and KamLAND. 

Table~\ref{tab:sensitivty} summarizes the detection ranges and alarm times, where the false alarm rate $F(t)$ is 1 yr$^{-1}$, for the four 15 $M_\odot$ models.
Roughly speaking, the detector sensitivity in SK-Gd and KamLAND 
\textcolor{black}{is similar}. 
Even so, KamLAND with $t_w=48$~hr has a better configuration for the alarm time.
\textcolor{black}{A} 35--45\% uncertainty of the detection ranges in pre-SN neutrino models \textcolor{black}{is seen}, even if their ZAMS mass is fixed to 15~$M_\odot$. 
The uncertainty of the alarm times is much larger. 
In the worst case, the alarm time at KamLAND has a range from 7.5--13~hr ($t_w=48$ hr, Patton model) to 17--26~hr ($t_w=48$ hr, Kato model) for the normal ordering.
The dependence of the mass ordering is also important. 
For the Kato and Yoshida models, it is difficult to realize an early alarm with the inverted ordering at SK-GD and KamLAND.
In contrast, KamLAND will be able to detect pre-SN signals 15--26~hr before collapse with $t_w=48$~hr in the normal ordering. 
Because the Odrzywolek and Patton models have similar numbers of DC events at KamLAND with $t_w=48$~hr, their detection ranges are also similar. 
The alarm times are, however, quite different; 11 -- 38~hr in the Odrzywolek model and 7.5 -- 13~hr in the Patton model.  
To discuss the detection time, we have to pay attention not only to the $\langle N_s \rangle$ itself but also to its time evolution.
\textcolor{black}{Among} all the models, JUNO has the best sensitivity to pre-SN neutrinos.
The detection range reaches 810 -- 1030~pc, and we \textcolor{black}{can} predict the Betelgeuse supernova 38 -- 74~hr prior to the collapse for the normal ordering with $t_w=48$ hr depending on models. 
Here, we estimate the detection ranges and the alarm times for $t<88$~hr \textcolor{black}{over the time window used for the Kato model calculation.}
Therefore, the sensitivity of JUNO should be improved by the longer calculations of pre-SN neutrinos. 
\textcolor{black}{Also, we use four 15~$M_\odot$ models in this study, assuming Betelgeuse-like targets. 
If we consider more massive stars, the detection range and alarm time will be enhanced~\cite{Asakura2016,simpson2019,Guo2019}.}

Finally, \textcolor{black}{it should be noted that we have not considered the actual detector responses (e.g., the energy resolution and quenching) in this work} because
these responses may not have made a significant difference in the previous studies of SK-Gd~\cite{simpson2019} and KamLAND~\cite{Asakura2016}.

\begin{table}[htbp]
\centering
\small
\caption{Detection ranges and alarm times for normal (inverted) mass ordering, where a false alarm rate is 1 yr$^{-1}$, for four pre-SN neutrino models with 15 $M_\odot$. }
  \begin{tabular}{l||c|cccc}\label{tab:sensitivty}
    Detector & Model & $N^{\textrm{DC}}_s(t=0.01)$   &Detection range [pc] & Alarm time [hr] &  $t_w$ [hr]\\
    \hline \hline
    \multirow{8}{*}{SK-Gd}& Kato &  \shortstack{46.7--49.9 (10.9--11.7)\\50.8--54.3 (12.2--13.0) \\ 54.3--58.0 (13.3--14.3) } & \shortstack{ 380--480 (180--230) \\ 350--460 (170--220) \\ 320--430 (160--210)} & 
    \shortstack{ 0.1-0.6 (--0.02) \\ 0.2--4.5 (--0.02) \\ 0.2--10 (--0.01)} &  \shortstack{12 \\ 24 \\ 48}\\ \cline{2-6}
    
    & Yoshida & \shortstack{21.4--22.8 (12.4--13.2)\\  26.3--28.0 (15.0--16.0)\\ 28.4--30.2 (16.1-17.2)} & \shortstack{ 260--330 (190--250) \\ 260--340 (190--260) \\ 240--320 (180--240)} & 
    \shortstack{ 0.1--1 (--0.1) \\ 0.4--6 (--0.2) \\ 0.2--6.5 (--0.2)} &  \shortstack{12 \\ 24 \\ 48}\\ \cline{2-6}
    
    & Odrzywolek &  \shortstack{45.3--48.3 (12.8--13.7)\\47.3--50.4 (13.4--14.3) \\49.1-52.4 (14.0-14.9)} & \shortstack{ 380--490 (200--260) \\ 340--460 (180--240) \\ 310--420 (170--220)} & \shortstack{ 4--6.5 (0.02--1.7) \\ 3--6.5 (--1.6) \\ 3--7 (--0.7)} &  \shortstack{12 \\ 24 \\ 48}\\ \cline{2-6}

    & Patton &  \shortstack{43.5--46.3 (12.9--13.9)\\ 45.8--48.9 (13.8--14.7)\\46.8--49.8 (14.1--15.0) } & \shortstack{ 370--480 (200--260) \\ 340--450 (180--250) \\ 310--410 (170--220)} & \shortstack{ 3.5--6 (0.02--0.9) \\ 3--6.5 (--0.5) \\ 2.5--5.5 (--0.1)} &  \shortstack{12 \\ 24 \\ 48}\\ \hline \hline

    \multirow{8}{*}{KamLAND}  & Kato &  \shortstack{7.6 (1.6)\\9.3 (2.1) \\ 10.9 (2.6)} & \shortstack{ 340--410 (150--190) \\ 350--440 (170--210) \\ 360--460  (180--220)}   &   \shortstack{ 0.2-1 (NA) \\ 5.5--20 (--0.02) \\ 17--26 (--0.1)} &  \shortstack{12 \\ 24 \\ 48}\\ \cline{2-6}
 
    & Yoshida &  \shortstack{4.5 (2.4)\\6.5 (3.5) \\7.7 (4.1)} &\shortstack{ 260--310 (190--230) \\ 290--370 (210--270) \\ 310--390  (220--280)}  &   \shortstack{ 0.5--16 (--0.1) \\ 8--18 (0.1--1.8) \\ 15--22 (0.3--7.5)} &  \shortstack{12 \\ 24 \\ 48}\\ \cline{2-6}

    & Odrzywolek &  \shortstack{9.7 (2.8)\\11.0 (3.1) \\12.4 (3.5)} &\shortstack{ 380--460 (200--240) \\ 380--480 (200--250) \\ 390--490  (200--260)} & \shortstack{ 5.5--8 (0.04--1.7) \\ 7--13 (0.08--2) \\ 11--38 (0.1--2.5)} &  \shortstack{12 \\ 24 \\ 48}\\ \cline{2-6}

    & Patton &   \shortstack{10.1 (2.9)\\11.4 (3.5) \\ 12.2 (3.6) } & \shortstack{ 390--470 (200--250) \\ 390--490 (210--260) \\ 380--490  (210--260)} & \shortstack{5.5--8.5 (0.07--1.9) \\ 7--11 (0.1--2.5) \\ 7.5--13 (0.1--3)}&  \shortstack{12 \\ 24 \\ 48}\\ \hline \hline

    \multirow{8}{*}{JUNO}  & Kato & \shortstack{232 (48.7) \\ 286 (65.2)\\ 341 (81.8)}&  \shortstack{950 (430) \\ 950 (440)\\ 960 (470)} & \shortstack{54 (24) \\ 64 (28) \\ 62 (34)}&  \shortstack{12 \\ 24 \\ 48}\\ \cline{2-6}
    
     & Yoshida &  \shortstack{142 (75.7) \\ 205 (109) \\247 (131) }& \shortstack{740 (540) \\ 810 (590) \\ 810 (590)} &  \shortstack{52 (30) \\ 64 (38) \\ 62 (46)} &  \shortstack{12 \\ 24 \\ 48} \\ \cline{2-6}
    
    & Odrzywolek & \shortstack{303 (86.2) \\344 (97.8) \\391 (111)} &   \shortstack{1090 (580) \\ 1050 (560) \\ 1030 (540)} & \shortstack{78 (14) \\ 76 (28) \\ 74 (48)} &  \shortstack{12 \\ 24 \\ 48}\\ \cline{2-6}
    
    & Patton &  \shortstack{315 (90.6) \\360 (106)\\ 385 (115)} & \shortstack{1110 (590) \\ 1070 (580) \\ 1020 (550)} & \shortstack{30 (17) \\ 34 (19) \\ 38 (20)} & \shortstack{12 \\ 24 \\ 48} \\

  \end{tabular}
\end{table}

\subsection{Early alarm with pre-supernova neutrino}
Recently, some projects attempt to realize an SN alarm with pre-SN neutrinos.
In this section, we give some examples of such activities.

\subsubsection{Alarm system at KamLAND}
KamLAND launched an alarm system with pre-SN neutrinos \textcolor{black}{in} 2015\footnote{https://www.awa.tohoku.ac.jp/kamland/SNmonitor/regist/index.html}. 
Signals \textcolor{black}{from photomultiplier tubes} are first digitized, and the energies of the events are reconstructed in the manner of the standard analysis at KamLAND.
It should be noted that KamLAND uses tentative calibration values and dead time tables in this process, 
for which a dedicated analysis is required.
Finally, \textcolor{black}{the} detection significance is calculated and released \textcolor{black}{online}.
The entire process takes $\sim$ 25~min, but it is possible to announce the detection in real time.
\textcolor{black}{Registration and authorization are required for access to the detection significance.}
The KamLAND collaboration is planning to implement an alarm distribution via the Gamma-ray Coordinates Network~(GCN)\footnote{https://gcn.gsfc.nasa.gov}.


\subsubsection{Alarm with SNEWS2.0}
Besides KamLAND, new detectors with high sensitivity to pre-SN neutrinos (e.g. SK-Gd, SNO+ and JUNO) \textcolor{black}{will be running} in the next few years.
An alarm system with multiple neutrino detectors will be realized soon, which will help prevent nearby SNe from escaping detection.
In addition, future detectors \textcolor{black}{will} have various detection channels, which \textcolor{black}{will} make it possible to detect all flavors of neutrinos and to perform multi-flavor studies for neutrino sources.
The ``SuperNova Early Waring System (SNEWS)'' is one  \textcolor{black}{such} candidate \cite{antonioli2004}. 
SNEWS involves an international collaboration representing current supernova neutrino-sensitive detectors~
(SK, LVD, IceCube, KamLAND, Borexino, DayaBay, and HALO).
It \textcolor{black}{aims to provide} the astronomy community with a prompt alarm for galactic supernovae, since neutrino signals emerge in the early stages of a supernova explosion.
\textcolor{black}{SNEWS is scheduled for an upgrade to SNEWS2.0, in which pre-SN neutrinos will be covered.}

%


\section{FINDINGS FROM FUTURE OBSERVATIONS} \label{sec4}
As discussed in the previous section, observations of pre-SN neutrinos have come into view.
\PRF{In this section, we then move on to the discussion about what we can learn from future observations of pre-SN neutrinos.}
We introduce three findings: distinction of progenitor models, restriction on convective properties, and determination of neutrino mass ordering.

  \begin{figure}
   \centering
   \includegraphics[scale=0.75]{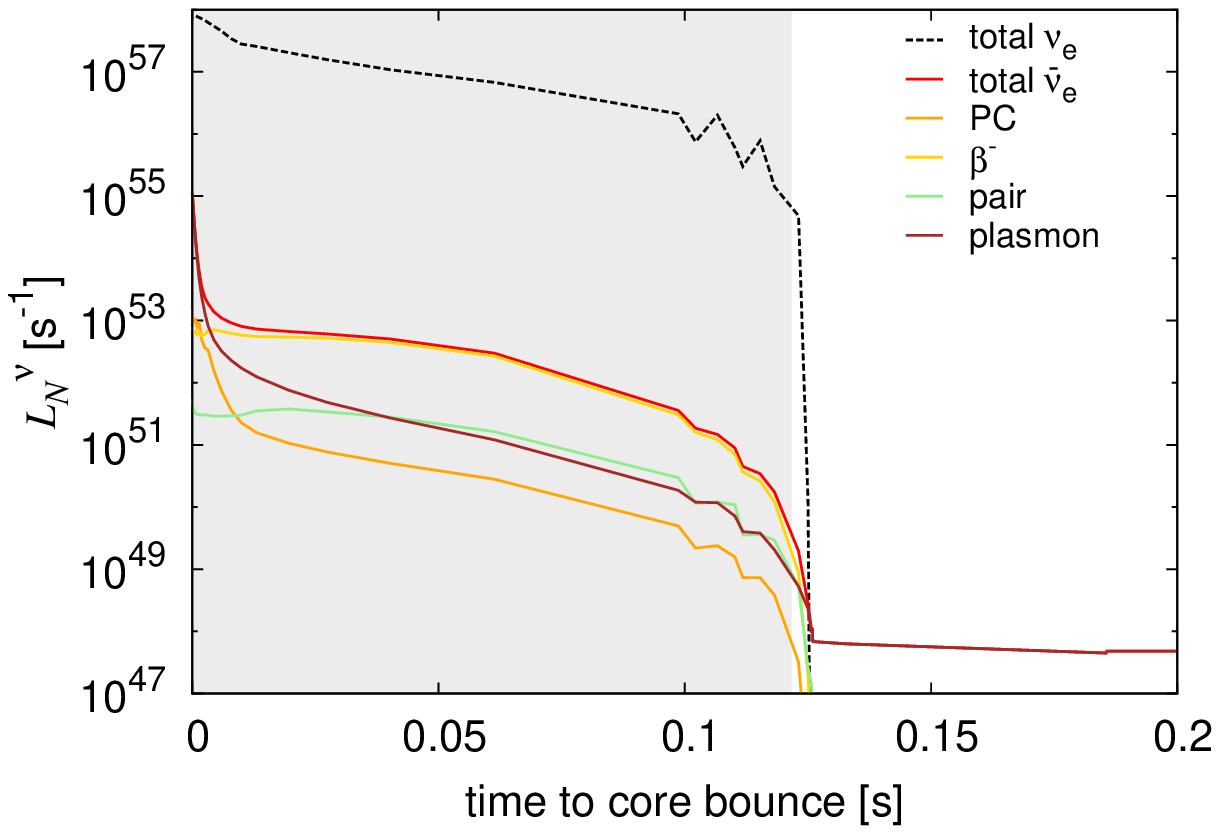}
   \caption{Time evolution of number luminosities for the  9~$M_\odot$ ECSNe-progenitor \cite{Kato2017}. The gray area represents the deflagration phase. Neutrino oscillations are not included.}
   \label{ref:onemg}
   \end{figure}

 \begin{table*}[htbp]
    \caption{Expected numbers of detection events until core bounce$^{\rm a,b}$. Numbers are pertinent to $\nu_e$'s for DUNE and to $\bar{\nu}_e$'s for JUNO. In the case of the 15~$M_\odot$ FeCCSN-progenitor, the individual contributions from the progenitor and collapse phases are also shown in that order in parentheses. The source is assumed to be located at $200\ {\rm pc}$ from the Earth. Data from Kato {\it et al.} \citep{Kato2017}.}
　   \label{eventrate}
    \begin{center}
    \begin{tabular}{ccccccccccc}
    \hline\hline
    detector & \multicolumn{2}{c}{9 $M_\odot$} & \multicolumn{2}{c}{15 $M_\odot$}  \\
    &normal & inverted & normal & inverted \\
    \hline
    JUNO    & 0.98 & 0.04  & 894 & 204\\
            &      &       & (891+3.07) & (203+0.63)\\
    DUNE    & 1765  & 22685  & 169 & 2142 \\
    (5MeV)  &      &       &(57.8+111)&(713+1429)\\ 
    DUNE    & 1238 & 15910 & 69.3 & 895 \\
    (10.8MeV)&    &    &    \\
    \hline
    \end{tabular}
    \end{center}
    \begin{tabnote}
    $^{\rm a}$Background noise was neglected in these estimations.\\
    $^{\rm b}$Table~\ref{tab:sensitivty} focuses on the time window for 0.1 hr before the core bounce; in this table, the event rates are integrated over a whole time range of progenitor and collapsing phases.
    \end{tabnote}
    \end{table*}

　\subsection{Distinction of progenitor models for core-collapse supernovae} 
    A CCSN's progenitor type is one of key determinants of its stellar evolution.
   There are two types of progenitors leading to ECSNe and FeCCSNe, as \PRF{discussed} in Section~\ref{stellarevo}.
   The boundary for the initial mass between \PRF{the} two progenitors is still unknown \cite{doherty2017,jones2013,Woosley2015}.
   The thermal evolutions of their cores are quite different because of the degree of degeneracy of electrons.
   In Figure~\ref{ref:rhoc_tc}, the 9~$M_\odot$ ECSN-progenitor is cooled via neutrino emission efficiently after carbon burning, and the central temperature becomes much lower than those in 12, 15 and 20~$M_\odot$ progenitors of FeCCSNe, for which the central temperatures increase continuously up to the onset of collapse. 
   The temperature rises rapidly in association with the O+Ne deflagration, which is a key feature \PRF{of} the ECSN-progenitor.
   
   Kato {\it et al.} \cite{Kato2015} employed state-of-the-art stellar evolution models and calculate pre-SN emissivity via pair annihilation and plasmon decay.
   In a follow-up article \cite{Kato2017}, they broadened the scope to include all flavors of neutrinos emitted from the pre-bounce phase.
   The time evolutions of \PRF{the} number luminosities for the 15~$M_\odot$ FeCCSN-progenitor and the 9~$M_\odot$ ECSN-progenitor are shown in Figures~\ref{ref:luminosity} and \ref{ref:onemg}, respectively.
   The number luminosity in the FeCCSNe-progenitor gradually increases from a few days before core bounce, whereas the number luminosity in the ECSNe-progenitor drastically increases at $t$ = 125~ms because of the O+Ne deflagration (see gray shaded region in Figure~\ref{ref:onemg}).
   The deflagration wave propagates outward to produce \PRF{the} NSE behind, and $\beta^-$ decay and ECs on heavy nuclei as well as on free protons efficiently emit $\bar{\nu}_e$'s and $\nu_e$'s.
   

   The expected numbers of neutrino events at JUNO and DUNE are summarized in Table~\ref{eventrate}, in which it is assumed that
   the progenitors are located at 200 pc.
   These estimations include neutrinos emitted in the collapse phase and neglect background noises \PRF{in} neutrino detectors discussed 
   in the previous section.
   It is found that JUNO can detect a few tens of $\bar{\nu}_e$'s from the FeCCSN-progenitor, even if they are emitted from 1 kpc away. 
   The detection of $\bar{\nu}_e$'s from the ECSN-progenitor seems to be nearly impossible, even with the planned detector. 
   We will \PRF{hence be} able to distinguish the two types of progenitors by detection or non-detection of $\bar{\nu}_e$'s.
   Kato {\it et al.} \cite{Kato2017} \PRF{also} estimated the expected numbers of $\nu_e$ events for the \PRF{two} values of \PRF{the} energy thresholds, 5 and 10.8 MeV, at DUNE.
   The energy of $\nu_e$'s in the collapse phase is high ($\sim$ 8 MeV), and we will still be able to detect a large number of $\nu_e$'s. 
   This implies that irrespective of the type and mass of \PRF{the} progenitor, we may be able to confirm our current understanding of the physics in the collapse phase.

    \subsection{Restriction on convective properties} \label{convection}
    Convections caused by the large energy generation \PRF{in} nuclear burnings affect the mixing of products and, in turn, change the core mass.
    \PRF{A} slight change \PRF{in mass} induces drastic change \PRF{in} the dynamics \PRF{of} the core, and convective properties are hence one of the key \PRF{phenomena} in the stellar evolution.
    \PRF{Our knowledge about these properties is, however, limited,} and
    their theoretical treatment \PRF{presents a large obstacle}. 
    It is possible that pre-SN neutrino observations may provide clues to help solve this problem \cite{Yoshida2016}.
    
\begin{figure}
    \centering
    \vspace*{-4.5cm}
    \includegraphics[scale=0.50,angle=270]{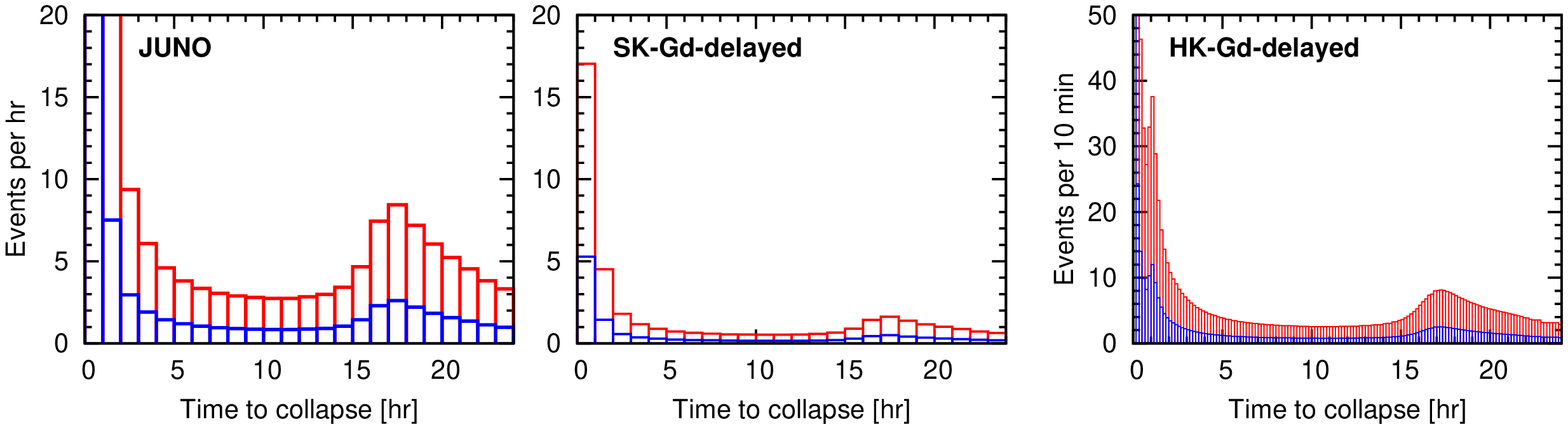}
    \vspace*{-4.5cm}
    \caption{Expected numbers of $\bar{\nu}_e$ events for the Yoshida model \citep{Yoshida2016} with 15 $M_\odot$ at JUNO, SK-Gd and HK-Gd from left to right.
    Events are measured per 1 hr for JUNO and SK-Gd and 10 min for HK-Gd.
    Red and blue bars indicate normal and inverted mass orderings, respectively.
    Detector parameters for JUNO and SK are shown in Table~\ref{tab:detector}.
    The detector efficiency for HK is assumed to be 0.5.
    }
    \label{ref:dndt_SKHK}
    \end{figure}

    Figure~\ref{ref:dndt_SKHK} shows the time evolution of neutrino events at JUNO (left), SK-Gd (middle) and HK-Gd (right).
    There are two peaks at $\sim$ 17 and 2 hr before core collapse, which correspond \PRF{respectively} to the ignitions of O-shell and Si-shell burnings. 
    The central temperature and density decrease because of the core expansion \PRF{at} the onsets of shell burnings,
    and these decreases lead to the suppression of neutrino emission in the core.
    \PRF{From the observational perspective,} $\bar{\nu}_e$'s emitted at the core have a larger contribution because they have high energies.
    The number of neutrino events \PRF{is hence} reduced because of shell burnings.
    Such features in neutrino events will confirm the existence of shell burnings, and provide information regarding this phenomenon (i.e.,  duration of shell burnings).
    
    \subsection{Determination of neutrino mass ordering}
    The observations of pre-SN neutrinos may help in determining the yet-known neutrino mass ordering.
    Because the behavior of the neutrino oscillation depends on the neutrino mass ordering, the number of neutrino events differ between the mass orderings.
    \PRF{In principle,} we will be able to detect more than a thousand $\nu_e$'s at DUNE, if the distance to the source is 200~pc and \PRF{if} neutrinos have the inverted mass ordering. 
    In contrast, \PRF{the} event numbers are reduced by a factor of $\sim$10 if they obey the normal mass ordering \cite{Kato2017}.

    Guo {\it et al.} \cite{Guo2019} suggested a model-independent method to determine the mass ordering using the numbers of IBD and ES events, $N_{\rm{IBD}}$ and $N_{\rm{ES}}$.
    Note that IBD is sensitive only to $\bar{\nu}_e$'s, whereas all flavors can be detected via ES.
    \PRF{The} $N_{\rm{ES}}/N_{\rm{IBD}}$ ratio \PRF{is} $\sim$ 0.91 and $\sim$ 3.8 at JUNO for the normal and inverted orderings, respectively, assuming pre-SN neutrinos are emitted from Betelgeuse.
    Guo {\it et al.} \PRF{found} that this large difference in $N_{\rm{ES}}/N_{\rm{IBD}}$ ratio between the normal and inverted orderings is insensitive to stellar models.
    These facts provide the basis for a model-independent determination of mass ordering.
    Although the current background rate for ES events is too high to perform this method, future reduction of backgrounds will make it possible.


\section{SUMMARY AND FUTURE PROSPECTS} \label{sec5}
\subsection{Summary}
Almost 30 years have passed since the historical supernova neutrino events of SN~1987A.
Remarkable progress has been made in neutrino detection techniques, opening neutrino astronomy to a new target, pre-SN neutrinos, which are mainly emitted from the cores of massive stars before core bounce.
This possibility was first pointed out by Odrzywolek {\it et al.} \cite{odrzywolek2004}, and followed by many works with state-of-the-art calculations of stellar evolutions.
Uncertainties in stellar physics will be clarified by pre-SN neutrino observations because neutrinos freely propagate through stars.
Their signals, moreover, issue a SN alarm for a few days before a supernova explosion and enable us to observe it.
In this review, we have introduced pre-SN neutrinos from both the theoretical and observational points of view.

Massive stars evolve by repeating the cycle of nuclear burning and gravitational core contraction until the ONe core is formed.
A stellar subsequent evolution depends on the ONe core mass; massive stars evolve into ECSNe or FeCCSNe depending on the initial mass.
Neutrinos have an important role in the thermal evolutions of the core.
In the case of the Kato model for the 15~$M_\odot$ FeCCSN-progenitor, the number luminosity of $\nu_e$'s gradually increases up to $L_N^\nu\sim10^{55}$ and $10^{57}\ \rm{s}^{-1}$ just before core collapse and core bounce, respectively.
ECs on heavy nuclei and free protons make large contributions to the total $\nu_e$ luminosity.
For $\bar{\nu}_e$'s, on the other hand, their number luminosity reaches a maximum $L_N^\nu\sim10^{53}\ \rm{s}^{-1}$ at the beginning of core collapse.
Neutrino emission via pair annihilation dominates the other reactions at $t>400$ s, whereas after this, $\beta^-$ decay overcomes it.
The average energies are $\sim$ 8 and 3 MeV for $\nu_e$'s and $\bar{\nu}_e$'s, respectively.
These properties of pre-SN neutrinos depend on stellar models and neutrino processes taken into account.
The peaks of number luminosities \PRF{vary in amplitude and width,} even if the initial mass is fixed and the investigation of the progenitor dependence is therefore strongly required.
In any case, pre-SN neutrinos are  smaller in number and have lower energies than do SN neutrinos, and detectors should have high sensitivity to low-energy neutrinos and a low background for the detection of pre-SN neutrinos.

\begin{figure}
   \centering
   \includegraphics[scale=0.8]{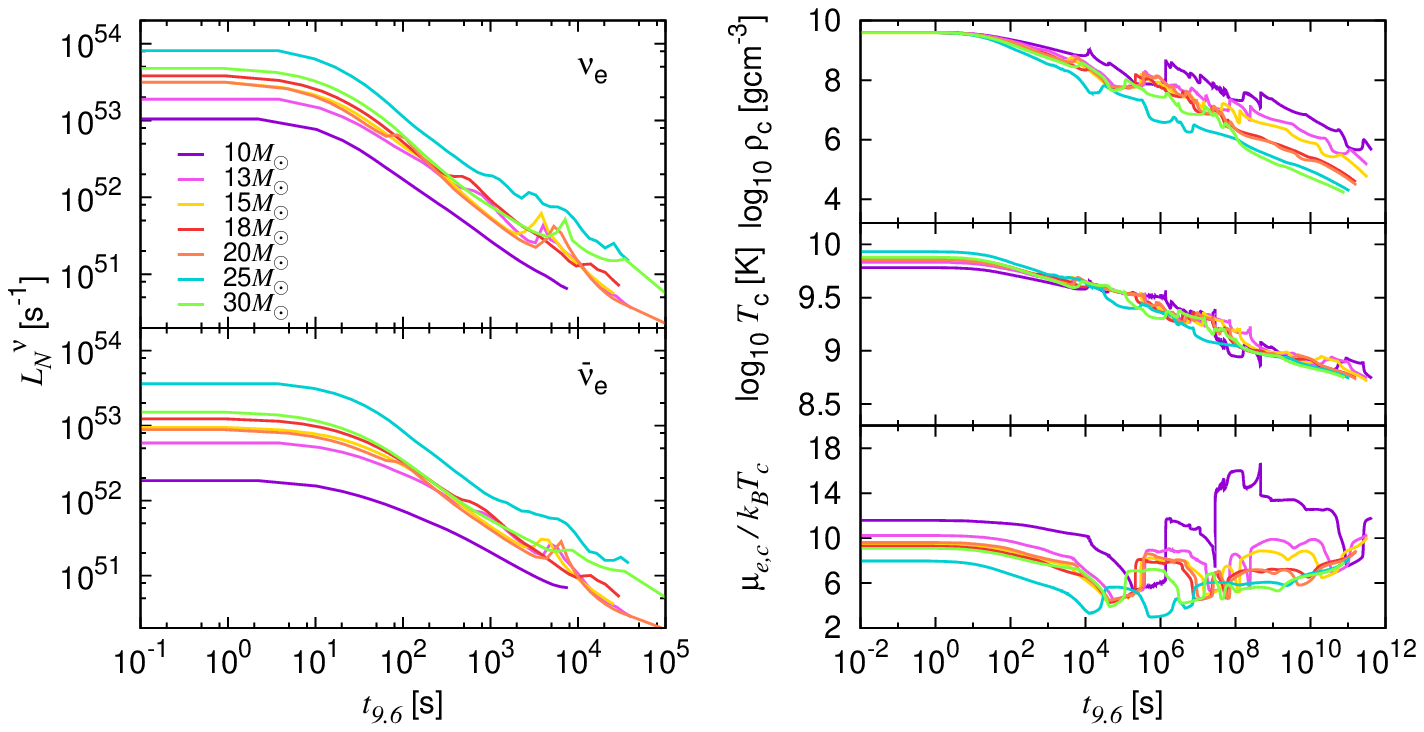}
   \caption{Left: Time evolutions of number luminosities of $\nu_e$'s (top) and $\bar{\nu}_e$'s (bottom) for seven progenitor models with different initial masses in Set ${\rm M_A}$ \cite{Yoshida2019}. Neutrino oscillations are not included. Right: Time evolutions of the central density, temperature, and degeneracy of electrons, where $\mu_{e,c}$ is the electron chemical potential at the center and $k_B$ is the Boltzmann constant. The origin of the time is defined when the central density becomes $\rho_c=10^{9.6}\ \rm{g~cm^{-3}}$.}
   \label{ref:systematic}
   \end{figure}

Various channels are currently employed in neutrino detection, and this variety makes it possible to detect multi-flavor neutrinos.
In detectors, considerable efforts have been made to reduce background noises, and the detection of pre-SN neutrinos come into view.
Detector sensitivity regarding pre-SN neutrinos has been discussed for different pre-SN models, detection statistics, and time windows.
We have hence presented a unified method for detector sensitivity with IBD on SK-Gd, KamLAND and JUNO and performed a comprehensive study using representative pre-SN neutrino models.
We have found that JUNO has the greatest sensitivity to pre-SN neutrinos and can detect pre-SN neutrinos from progenitors even at $\sim$ 1~kpc.
SK and KamLAND, on the other hand, have a similar sensitivity and will predict the explosion of Betelgeuse at 200 pc.
We have also discussed an SN alarm for the same target and have found that JUNO will be able to issue an alarm $\sim$ 78 hr prior to collapse.
The detection range and alert time depend on a time window $t_w$ for the cumulative number of events and pre-SN neutrino models.
The most promising $t_w$ is determined by keeping a balance between the increments of $\langle N_s \rangle$ and $\langle N_g \rangle$ with $t_w$.
The longer $t_w$ provides better sensitivity at KamLAND because of the low background rates, whereas the sensitivity of SK reaches its maximum with the shortest $t_w$. 
JUNO has intermediate dependence between those of SK-Gd and KamLAND. 
There is 35 -- 45\% uncertainty of the detection range in the pre-SN neutrino models, even if their ZAMS mass is fixed at 15 $M_\odot$.  
The alarm times have a larger uncertainty.
In the worst case for the normal ordering, the alarm time at KamLAND has a range from 7.5--13~hr ($t_w=48$ hr, Patton model) to 17--26~hr ($t_w=48$ hr, Kato model).

   \begin{table*}[htbp]
    \small
    \caption{RSGs within 1 kpc from the Earth. Data from Nakamura {\it et al.} \cite{nakamura2016} and reference therein. \label{candidate}}
    \begin{tabular}{cc|cc|cc|ccccc}
    \hline\hline
    RSG & distance (pc) & RSG & distance (pc) & RSG & distance (pc) & RSG & distance (pc)\\
    \hline
    $\epsilon$ Peg & 150 & $h^1$ Pup & 420 & V424 Lac & 600 & HD203338 & 870 \\
    $\alpha$ Sco & 160 & 145 CMa & 440 & HD217694 & 600 &  SW Cep & 870 \\
    $\lambda$ Vel & 190 &  1 Pup & 440 & ES And & 600 & V557 Cep & 870 \\
    $\zeta$ Cep & 200 & $\sigma$ CMa & 480 & $\psi^1$ Aur & 710 & $\mu$ Cep & 870 \\
    $\alpha$ Ori & 220 & $o^1$ CMa & 480 & 41 Gem & 710 & RT Cep & 870 \\
    q Car & 360 & BM Sco & 480 &  HR5742 & 710 & VV Cep & 870 \\
    119 Tau & 370 & HR3692 & 540 & HR8248 & 790 & AZ Cep & 870 \\
    w Car & 390 & 12 Peg & 600 & ST Cep & 790 & CK Cep & 870 \\
    47 Cyg & 390 & 5 Lac & 600 &  $\theta$ Del & 860 & V809 Cas & 870 \\
    $\xi$ Cyg & 390 & V418 Lac & 600 &  V419 Cep & 870 & MY Gem & 1000 \\
    &  &  &  &  &    & HR861  & 1000 \\
    \hline
    \end{tabular}
    \end{table*}

The importance of pre-SN neutrinos has been discussed in three points so far: distinction of progenitor models for CCSNe, restriction on convective properties, and determination of neutrino mass ordering.
Firstly, Kato {\it et al.} \cite{Kato2015, Kato2017} found that we can distinguish two types of progenitors in CCSNe---FeCCSNe and ECSNe---by the detection or non-detection of $\bar{\nu}_e$'s.
The thermal evolutions of these progenitors are quite different in the late phase because of the difference in the degeneracy of electrons, and the distinction of progenitors is important in establishing the stellar evolution theory.
Secondly, shell burning, another important phenomenon, is also found in neutrino signals.
Yoshida {\it et al.} \cite{Yoshida2016} found two peaks in the time evolution of neutrino signals due to the suppression of neutrino emission by the onsets of O-shell and Si-shell burnings.
The pre-SN neutrino observations will provide information about the properties of shell burnings (i.e., duration).
Finally, observations of pre-SN neutrinos may help in determining the yet-known neutrino mass ordering.
The neutrino fluxes at the Earth depend on the mass ordering because of neutrino oscillations, as shown by the difference ratios in the number of ES to IBD events.
The $N_{\rm{ES}}/N_{\rm{IBD}}$ ratio is $\sim$ 0.91 and 3.8 at JUNO for the normal and inverted orderings, respectively, assuming pre-SN neutrinos are emitted from Betelgeuse.
Such a large difference between mass orderings is model-independent and future measurements of this ratio will hence reveal the mass ordering.

\subsection{Future prospects}
Theoretical studies of pre-SN neutrinos have just started, and much remains to be done for the future pre-SN neutrino observations.
Systematical studies \PRF{using progenitors with} various initial masses are the first priority.
Within the detectable range for pre-SN neutrinos $<$ 1 kpc from the Earth (see Table~\ref{tab:sensitivty}), there are 41 candidate RSGs, which are summarized in Table~\ref{candidate} (see Nakamura {\it et al.} \cite{nakamura2016} and references therein).
Their initial masses are expected to be within a range of 9 -- 25 $M_\odot$, although it is difficult to determine them from observations.
The evolution of a massive star is highly dependent on its initial masses (see the right panels of Figure~\ref{ref:systematic}) and we should hence take this dependence into account in the prediction of pre-SN neutrinos.
Actually, we find that the number luminosities in both flavors differ by more than a factor of 10 with the initial mass.
The left panels of Figure~\ref{ref:systematic} show the number luminosities of $\nu_e$'s (top) and $\bar{\nu}_e$'s (bottom).
We also find that the number luminosities have positive and negative correlations with the central temperature and degeneracy of electrons in the middle and bottom panels on the \textcolor{black}{right}.
These correlations may provide clues for stellar physics in the future.
We should proceed with such theoretical studies to get precise information from the future observations.

From the observational point of view, the first priority is to establish an alarm system with multiple neutrino detectors.
Detection at multi-detectors makes observational data reliable and more precisely imposes observational constraints on the stellar evolution and the explosion mechanism.
In particular, the data derived from the detectors with sensitivity to different neutrino flavors broaden our findings.  
We may, moreover, point to neutrino sources using the neutrino detectors at different locations \cite{linzer2019}.
Hence, there is a strong need to establish a combined-alarm system with multi-detectors such, as the SNEWS project.




\section*{ACKNOWLEDGMENTS}
The writing of this review was supported by the Grant-in-Aid for Scientific Research on Innovative Areas (No. 26104007, 19H05803) from the Ministry of Education, Culture, Sports, Science and Technology and the Grant-in-Aid for the Scientific Research of Japan Society for the Promotion of Science (No. JP17H01130).
C.K. is supported by Tohoku University Center for Gender Equality Promotion (TUMUG) Support Project (Project to Promote Gender Equality and Female
Researchers).
We would like to thank Editage (www.editage.com) for English language editing.

%

\noindent

\bibliographystyle{ar-style5.bst}
\bibliography{ref.bib}

\end{document}